\def\countMode{0}

\def\hlMode{1}

\def\anoMode{0}

\def\noteMode{1}

\documentclass[%
 aip,
 amsmath,amssymb,
 reprint,%
]{revtex4-1}

\usepackage{graphicx}% Include figure files
\usepackage{dcolumn}% Align table columns on decimal point
\usepackage{bm}% bold math
\usepackage{color,soul}

\usepackage[utf8]{inputenc}
\usepackage[T1]{fontenc}
\usepackage{mathptmx}
\usepackage{etoolbox}
\usepackage{hyperref}
\usepackage{cleveref}
\usepackage{cases}
\usepackage{mathtools}
\usepackage{stackengine}
\stackMath

\DeclareMathOperator{\sinc}{sinc}

\newcommand{\U}[1]{\mathrm{\; #1}}
\newcommand{\E}[1]{\cdot 10^{#1}}
\newcommand{\Eb}[1]{10^{#1}}
\newcommand{\hel}{\psi}

\newcommand{\ed}[1]{\E{#1}\U{m^{-3}}}
\newcommand{\ebd}[1]{\Eb{#1}\U{m^{-3}}}

\newcommand{\lb}{\left(}
\newcommand{\rb}{\right)}
\newcommand{\br}[1]{\lb #1 \rb} % regular ( )
\newcommand{\sbr}[1]{\left[ #1 \right]} % square [ ]
\newcommand{\pbr}[1]{\left\{ #1 \right\}} % pointy  { }

\renewcommand{\vec}[1]{\bm{#1}}

\newcommand{\vE}{\vec{E}}

\newcommand{\vj}{\vec{j}}

\newcommand{\uniVec}[1]{\vec{\hat{#1}}}
\newcommand{\uniZ}{\uniVec{z}}
\newcommand{\uniR}{\uniVec{r}}
\newcommand{\uniPhi}{\uniVec{\phi}}

\newcommand{\w}{\omega}
\newcommand{\dens}{m^{-3}}
\newcommand{\dg}{^\circ}   % degree symbol

\newcommand{\rect}[1]{\Pi\left( #1 \right)}

\newcommand\tsup[2][2]{%
 \def\useanchorwidth{T}%
  \ifnum#1>1%
    \stackon[-.5pt]{\tsup[\numexpr#1-1\relax]{#2}}{\scriptscriptstyle\sim}%
  \else%
    \stackon[.5pt]{#2}{\scriptscriptstyle\sim}%
  \fi%
}

\newcommand{\ft}[1]{\mathcal{F}\left[#1\right]}

\newcommand{\ftFac}[1]{e^{- i #1}}
\newcommand{\iftFac}[1]{e^{ i #1}}
\newcommand{\ftSinc}[1]{\sinc\left(\frac{#1}{2\pi}\right)}
\newcommand{\sft}[1]{\tsup[1]{#1}}

\newcommand{\expM}[2][-]{e^{#1#2 i m \pi}}
\newcommand{\expA}[1][-]{e^{#1 i m\alpha z}}
\newcommand{\expB}[1][-]{e^{#1 i m \beta}}

\newcommand{\conj}[1]{\overline{#1}}

\newcommand{\avg}[1]{\left\langle #1 \right\rangle}

\newcommand{\dsft}[1]{\tsup[2]{#1}}

\newcommand{\ld}{lead}
\newcommand{\eq}{equal}
\newcommand{\su}{supporting}

\usepackage{ifthen}
\newcommand{\optCredit}[3][;]{
\ifthenelse{\equal{#3}{}}{}{#2 (#3)#1}}

\newcommand{\hlcolor}[2]{\sethlcolor{#1}\hl{#2}}

\if\noteMode1
    \renewcommand{\hlcolor}[2]{}
\fi

\if\countMode1
    \renewcommand{\cite}[1]{{}}
\fi

\if\hlMode0
    \renewcommand{\hlcolor}[2]{#2}
    
\fi

\if\hlMode1
    \let\oldcite\cite
    \renewcommand{\cite}[1]{\mbox{\oldcite{#1}}}
\fi

\newcommand{\ano}[2][]{#2}
\if\anoMode1
    \renewcommand{\ano}[2][]{#1}
\fi

\newcommand{\MAPL}{\ano[MAP experiment ]{\textit{Madison AWAKE Prototype (MAP)} }} 

\newcommand{\MAPLC}{\ano[MAP Experiment ]{\textit{Madison AWAKE Prototype (MAP)} }} 

\makeatletter
\def\@email#1#2{%
 \endgroup
 \patchcmd{\titleblock@produce}
  {\frontmatter@RRAPformat}
  {\frontmatter@RRAPformat{\produce@RRAP{*#1\href{mailto:#2}{#2}}}\frontmatter@RRAPformat}
  {}{}
}%
\makeatother

\begin{document}

\title{Computational and Analytical Optimization of Helicon Antennas with a Fast Full Wave Solver Exploiting Azimuthal Fourier Decomposition}
% Force line breaks with \\
\ano{
\author{Marcel Granetzny}
 \email{granetzny@wisc.edu}
\author{Oliver Schmitz}
\affiliation{University of Wisconsin - Madison, Department of Nuclear Engineering and Engineering Physics, Madison, WI 53706, USA}
}

\date{\today}% It is always \today, today,

\begin{abstract}
Plasma wakefield accelerators (PWA), such as \textit{AWAKE}, require homogenous high-density plasmas. The \MAPL has been built to create a uniform argon plasma in the $\Eb{20}\U{\dens}$ density range using helicon waves. Computational optimization of MAP plasmas requires calculating the helicon wavefields and power deposition. This task is computationally expensive due to the geometry of high-performance half-helical antennas and the small wavelengths involved. We show here for the first time how the 3D wavefields can be accurately calculated from a small number of 2D-axisymmetric simulations. Our approach exploits an azimuthal Fourier decomposition of the non-axisymmetric antenna currents to massively reduce computational cost and is implemented in the Comsol finite-element framework. This new tool allows us to calculate the power deposition profiles for 800 combinations of plasma density, antenna length and radial density profile shape. The results show the existence of an optimally coupling antenna length in dependence on the plasma density. This finding is independent of the exact radial profile shape. We are able to explain this relationship physically through a comparison of the antenna power spectrum with the helicon dispersion relation. The result is a simple analytical expression that enables power coupling and density optimization in any linear helicon device by means of antenna length shaping.  
\end{abstract}

% Use showkeys class option if keyword display is desired
\keywords{helicon, plasma waves, full-wave simulation, antennas, power coupling, optimization, half-helical}

\maketitle
\preprint{AIP/123-QED}

% \header{Introduction}
\section{Introduction}

Helicon waves\cite{Boswell1984,Chen1991} are magnetized plasma waves with a wide range of applications, including semiconductor manufacturing\cite{Tynan1997}, nuclear fusion\cite{Rapp2016,Rapp2017}, space propulsion\cite{Squire2006} and plasma-based particle acceleration\cite{Buttenschon2018,gschwendtner2016awake}. Their wave fields define the local heating power deposition, thereby shaping the plasma density, temperature, and neutral profiles. These profiles in turn guide the wavefield and power deposition patterns, resulting in a self-consistent plasma equilibrium between RF input power, ionization, various particle and energy-loss channels, and plasma transport. A detailed understanding of the RF power deposition in a helicon plasma is therefore the prerequisite to tailor the plasma profiles to a given application.\\

To this end, we have developed a new 3D wavefield solver in Comsol\cite{ComsolIntro}, a widely used finite element framework. The solver calculates the helicon wavefields and power deposition in a given plasma and neutral background using the cold-plasma wave formalism\cite{Stix1992} and exploits the cylindrical geometry of helicon discharges to simplify the problem to a small number of 2D-axisymmetric simulations. While a similar approach has been used in the past\cite{Piotrowicz2018}, previous implementations could not account for the 3D structure of the helical antennas used in high-performance experiments. We solve this problem through an azimuthal Fourier decomposition of the antenna currents. The result is a full-wave solver that runs on an industry-proven finite element framework and can compute the wavefields at a massively reduced cost compared to direct 3D computation.\\

We leverage this new capability to optimize RF power-coupling into the plasma. The motivation driving this work is to find a configuration that minimizes input power while satisfying the plasma density and homogeneity requirements for the \textit{AWAKE} plasma wakefield accelerator project\cite{Assmann2014,Muggli2018}. By varying the antenna geometry for a wide range of plasma densities, we find the optimal antenna length for plasmas covering two orders of magnitude in density space. We are then able to reproduce these findings analytically starting from first principles.\\

In the rest of this introduction, we will briefly introduce the \textit{AWAKE} project and the \MAPL. We will then analyze the helicon dispersion relation and derive requirements for the needed finite element size. We develop our computational framework in \cref{sec:Methods}. \Cref{sec:Results} shows our main results, namely the identification of dominant modes, and comparison to analytical expectations in a uniform plasma and power coupling studies. We find an intuitive and analytical model for the observed power coupling optima in \cref{sec:Discussion} and summarize our work in \cref{sec:Summary}.

\subsection{Helicon Plasmas for the AWAKE Project}

The \textit{Advanced Proton Driven Plasma Wakefield Acceleration Experiment (AWAKE)} at CERN\cite{gschwendtner2016awake} has been built to develop the technology for practical beam-driven wakefield acceleration of electrons.\cite{Tajima1979,malka2013review,joshi2004review,cakir2019brief} Electron and positron accelerators are currently limited to the low hundreds of GeV range due to synchrotron losses in even the largest accelerators such as the Large Electron-Positron Collider (LEP).\cite{ABBIENDI2002233} The first proof of concept application will be a hundred-meter range accelerator, driven by proton bunches from the Super Proton Synchrotron (SPS). In a later stage protons from the Large Hadron Collider (LHC) would be used as drivers for the world's first $\U{TeV}$ lepton collider with energy gains in the $\U{GeV/m}$ range.\cite{caldwell2011plasma,Caldwell_2016}\\

Plasma wakefield acceleration relies on highly uniform high-density plasmas. AWAKE requires a plasma density of $\Eb{20}\U{\dens}$ and on-axis density-uniformity of ideally $0.25$\% to achieve an energy gain of $1$ GeV/m\cite{Muggli2018}. Achieving these density parameters in a reproducible, cost-efficient, and scalable way is challenging. One of the most promising avenues is plasma breakdown and sustainment by helicon waves. Densities in the mid $\Eb{20}\U{\dens}$ range have been achieved in high-power helicon plasmas.\cite{Buttenschon2018} However, these plasmas have significant axial density variations thus requiring significant advances in density profile optimization.

\subsection{The \MAPLC}

\begin{figure*}
\includegraphics[width=0.7\linewidth]{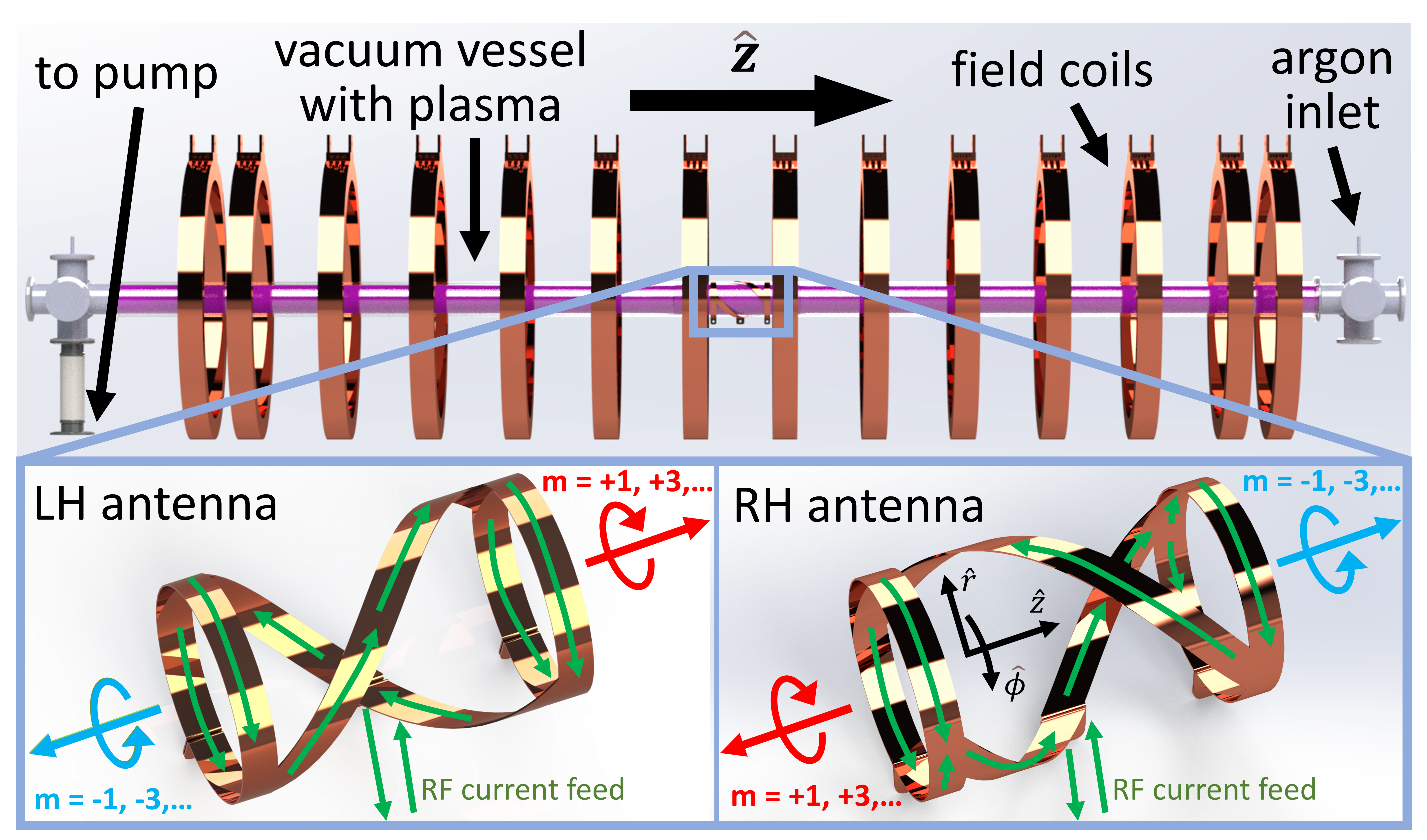}
			\caption{\label{fig:MAPCAD}CAD model of the core components of the \MAPL. Shown are the magnetic field coils, vacuum vessel with argon plasma, and helicon antenna (center). The antennas used in our simulations are either left-helical (LH) or right-helical (RH). Both versions are shown at the bottom along with the spatial rotation and propagation directions of the different azimuthal modes denoted by wave number $m$. Modes that rotate in a right-handed (red) or left-handed (blue) sense around $\uniZ$ are launched in opposite directions. These launch directions are reversed for opposite antenna helicities. The direction of current flow during one half of the RF cycle is indicated in green.}
\end{figure*}

The \MAPL is a new \ano[helicon plasma source]{experiment at the University of Wisconsin - Madison}. MAP has been built as a dedicated plasma source development platform for the AWAKE project. A more detailed description of the experiment will be published elsewhere, but the core geometry is shown in \cref{fig:MAPCAD}.\\

MAP uses up to 20 kW of RF power, spread over two antennas at 13.56 MHz in a 50 mT background field, to sustain a helicon argon plasma inside a borosilicate vacuum vessel with an inner diameter of 52 mm and a total length of 2.6 m. MAP has previously been used to reveal for the first time the mechanism behind the preference of right-handed modes and discharge directionality in helicon plasmas\cite{Granetzny_2023}. A recent study has used MAP to perform the first measurement of the 2D ionization source rate in a helicon device\cite{Zepp_2024}. The simulations shown in this study were all performed using the MAP geometry and magnetic field.\\

\subsection{Helicon Dispersion Relation}

Helicon waves are in essence bounded whistler waves\cite{Chen2015} and their dispersion can be derived directly from the cold plasma dielectric tensor\cite{Stix1992}. For a uniform plasma with density $n_e$, in a magnetic field of strength $B$, the dispersion relation at RF frequency $f$ in cylindrical coordinates is\cite{Chen1997}

\begin{align}
\label{eq:heliconDispRel}
0 &= \delta \beta^2 - k \beta + k_w^2\\
\label{eq:kw}
k_w^2 &= \frac{2\pi f n_e \mu_0 e}{B}\\
\label{eq:delta}
\delta &= \br{2\pi f + i \nu}\frac{m_e}{e B},
\end{align}

where $\beta$ and $k$ are the total and axial wavenumbers, respectively, and collisional damping is accounted for by the effective combined electron-ion and electron-neutral collision frequency $\nu$.\\

The total wave number $\beta$ is related to the radial and axial wave numbers, $T$ and $k$, respectively, as\cite{Woods1962}

\begin{align}
\label{eq:waveNumRels}
\beta^2 &= T^2 + k^2.
\end{align}

In addition, in a cylindrical plasma, the helicon wave splits into discrete azimuthal modes, designated by wavenumber $m$.\\

\begin{figure}
\includegraphics[width=\linewidth]{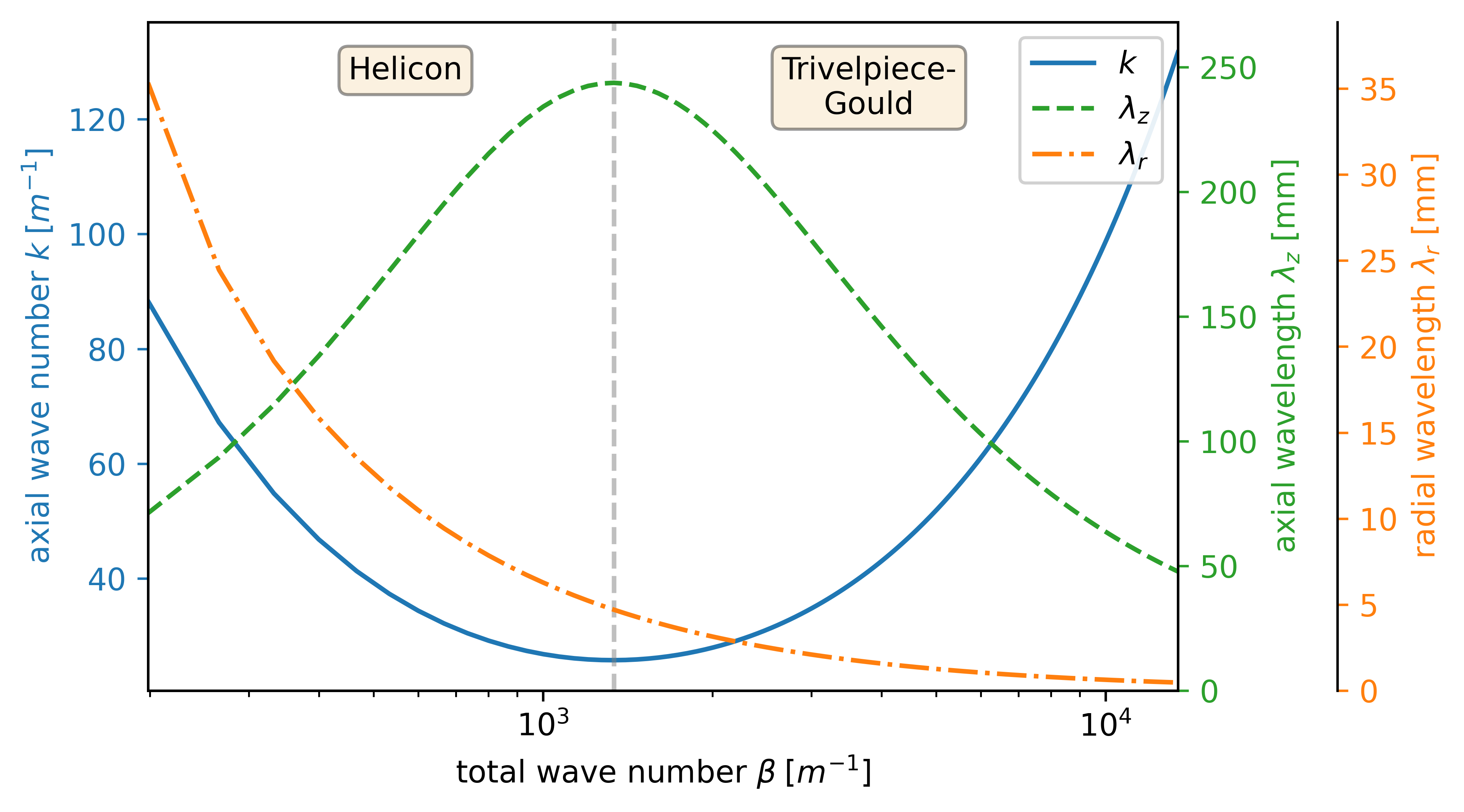}
			\caption{\label{fig:kBetaPlot}Helicon-TG dispersion relation with corresponding axial and radial wave lengths in a uniform plasma with $B = 50 \U{mT}$, $n_e = 5\E{19}\U{\dens}$ at $f = 13.56 \U{MHz}$. The dispersion relation splits into the helicon and Trivelpiece-Gould (TG) branches at the minimum axial wave number. Axial wavelengths are an order of magnitude larger than radial wavelengths over most of the dispersion relation and up to two orders of magnitudes larger in the high $\beta$ parts of the dispersion relation.}
\end{figure}

The dispersion relation and corresponding axial and radial wavelengths for a medium density ($5\E{19}\U{\dens}$) plasma at moderate field (50 mT) at 13.56 MHz are shown in \cref{fig:kBetaPlot}. For each axial wavenumber $k$ there are two solutions, the helicon and Trivelpiece-Gould (TG) branches, with the helicon branch on the left and the TG branch on the right in \cref{fig:kBetaPlot}. Radial wavelengths range from a few millimeters in the TG branch to tens of millimeters in the helicon branch. In contrast, axial wavelengths are higher but span a narrower range from tens to hundreds of millimeters. In consequence, the helicon branch propagates predominantly at a shallow angle to the z-axis, whereas the TG branch propagates almost perpendicular to it.\\

Looking at the smallest radial wavelength as compared to the largest axial wavelength, we find that they differ by about two orders of magnitude. The small radial wavelengths combined with the scale difference between the radial and axial directions make it computationally expensive to simulate these wave fields. Since the TG mode is strongly damped, it is often excluded from calculations to avoid this difficulty. However, in practice, TG mode damping leads to a strong power deposition at the edge, which is important for igniting and maintaining the plasma, especially at higher densities.\cite{Chen2015,Curreli2011}\\

\section{Methods}
\label{sec:Methods}

Helicon plasmas are typically created in an axisymmetric configuration, such as the one in \cref{fig:MAPCAD}, and as a result, have axisymmetric plasma profiles\cite{Zepp_2024}. This symmetry can be exploited to massively reduce computational cost by performing a 2D-axisymmetric simulation instead of solving the full 3D problem. However, the best-performing helicon experiments use half-helical antennas\cite{Sudit1996,Chen2015}  of the type shown in the bottom of \cref{fig:MAPCAD}. The 3D nature of these antennas breaks axisymmetry and traditionally makes a full 3D calculation necessary. Since the wavelength, and therefore the necessary mesh element size, decreases with increasing density, these simulations become computationally expensive at high density. A 3D simulation of a moderate-size plasma then requires employing a cluster computer, simulating only low densities or neglecting small wavelengths and therefore the TG mode which is responsible for most of the power deposition\cite{Chen2015}. \\

However, it is well known that a small number of azimuthal modes are responsible for the entirety of wavefields measured in helicon discharges.\cite{Chen1996a} Our strategy going forward is to split the antenna currents into discrete azimuthal modes. We can then solve the 2D-axisymmetric problem for every azimuthal mode of interest and combine them to reconstruct the full 3D solution from a small number of 2D-axisymmetric simulations.

\subsection{Azimuthal Fourier Decomposition of Antenna Currents}
\label{sec:AziFT}

The lower half of \cref{fig:MAPCAD} shows examples of half-helical antennas used on MAP and other high-density helicon experiments. Overlaid on the structure of the antennas itself are green arrows indicating the direction of the currents during one half of the RF cycle. In addition, we show the cylindrical coordinate system using $\uniR$, $\uniPhi$, $\uniZ$ notation.\\

\begin{figure}
\includegraphics[width=\linewidth]{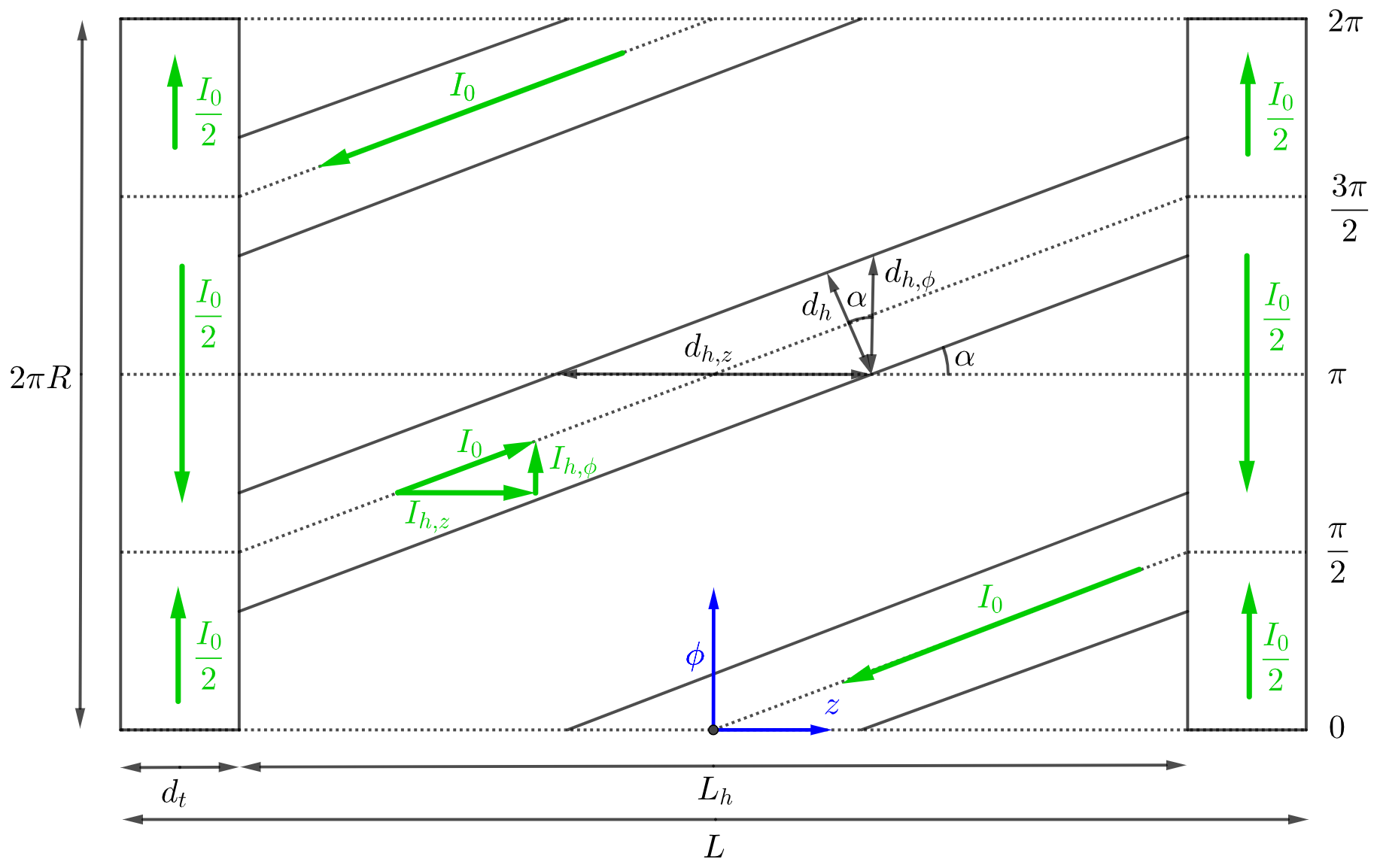}
			\caption{\label{fig:antennaGeoMain}Flattened geometry of a right-handed half-helical antenna of the type shown on the bottom right in \cref{fig:MAPCAD}. RF current directions and magnitudes are shown in green, coordinate system directions and origin in blue, and general antenna shape and dimensions in black.}
\end{figure}

Since the skin effect will limit currents to a very thin layer on the inner side of the antenna we can model the currents on it as surface currents, flowing only in the $\uniPhi$ and $\uniZ$ directions. \Cref{fig:antennaGeoMain} shows the currents of a right-helical antenna (lower right in \cref{fig:MAPCAD}) in the $(\phi, z)$ plane. Here $L$ is the overall length of the antenna, $d_t$ is the width of the transverse straps or end hoops, and $d_h$ is the width of the helical antenna straps. $L_h$ is the axial length of the helical part of the antenna.\\

A geometric analysis of \cref{fig:antennaGeoMain}, shown in \cref{app:antCurReal}, establishes that the surface current components in the azimuthal ($K_\phi$) and axial ($K_z$) directions for an antenna centered at axial position $z=0$ are

\begin{align}
\label{eq:Kz}
K_z =& \frac{I_0 \psi}{d_h \sqrt{1+\gamma^2}}\sum^{2}_{n=0} \sbr{ (-1)^{n+1}\rect{\frac{\phi- n\pi -\frac{\pi \hel  z}{L_h}  }{\phi_w}} }\nonumber\\
&\times\rect{\frac{\phi - \pi}{2\pi}}\rect{\frac{z}{L_h}}\\
\label{eq:Kphi}
K_{\phi} =& \frac{I_0}{2 d_t}\left[\rect{\frac{\phi-\pi}{2\pi}} -2\rect{\frac{\phi-\pi}{\pi}}\right]\nonumber\\
&\times\left[\rect{\frac{z-z_L}{d_t}}+\rect{\frac{z-z_R}{d_t}}\right] + \frac{\gamma}{\hel}K_z,
\end{align}
% \end{widetext}

where $I_0$ is the total antenna current and the following conventions are used

\begin{align}
\label{eq:LhDef}
L_h &= L - 2 d_t\\
\gamma &= \frac{\pi R}{L_h}\\
\phi_w &= \sqrt{1+\gamma^2}\frac{d_h}{R}\\
\label{eq:antCurSumEnd}
z_{L/R} &= \mp \frac{L-d_t}{2}\\
\label{eq:psi}
\hel &= 
\begin{cases}
+1 \quad\text{for a right-helical antenna}\\
-1 \quad\text{for a left-helical antenna}\\
\end{cases}\\
\label{eq:rect}
\rect{x} &= 
\begin{cases}
1 & \text{if}  \quad -0.5 < x < 0.5 \\
0 & \text{else}
\end{cases}
\end{align}

To perform the azimuthal mode decomposition of the antenna currents, we use the following Fourier series definitions for azimuthal angle $\phi$ and azimuthal wavenumber $m$

\begin{align}
\label{eq:mphiFtDef2}
f(\phi) &= \sum_{m} \sft{f}(m) \iftFac{m \phi},\\ 
\label{eq:mphiFtDef}
\sft{f}(m) &= \frac{1}{2\pi} \int_{0}^{2\pi} f(\phi) \ftFac{m \phi} \,d\phi . 
\end{align}

\begin{figure}
\includegraphics[width=\linewidth]{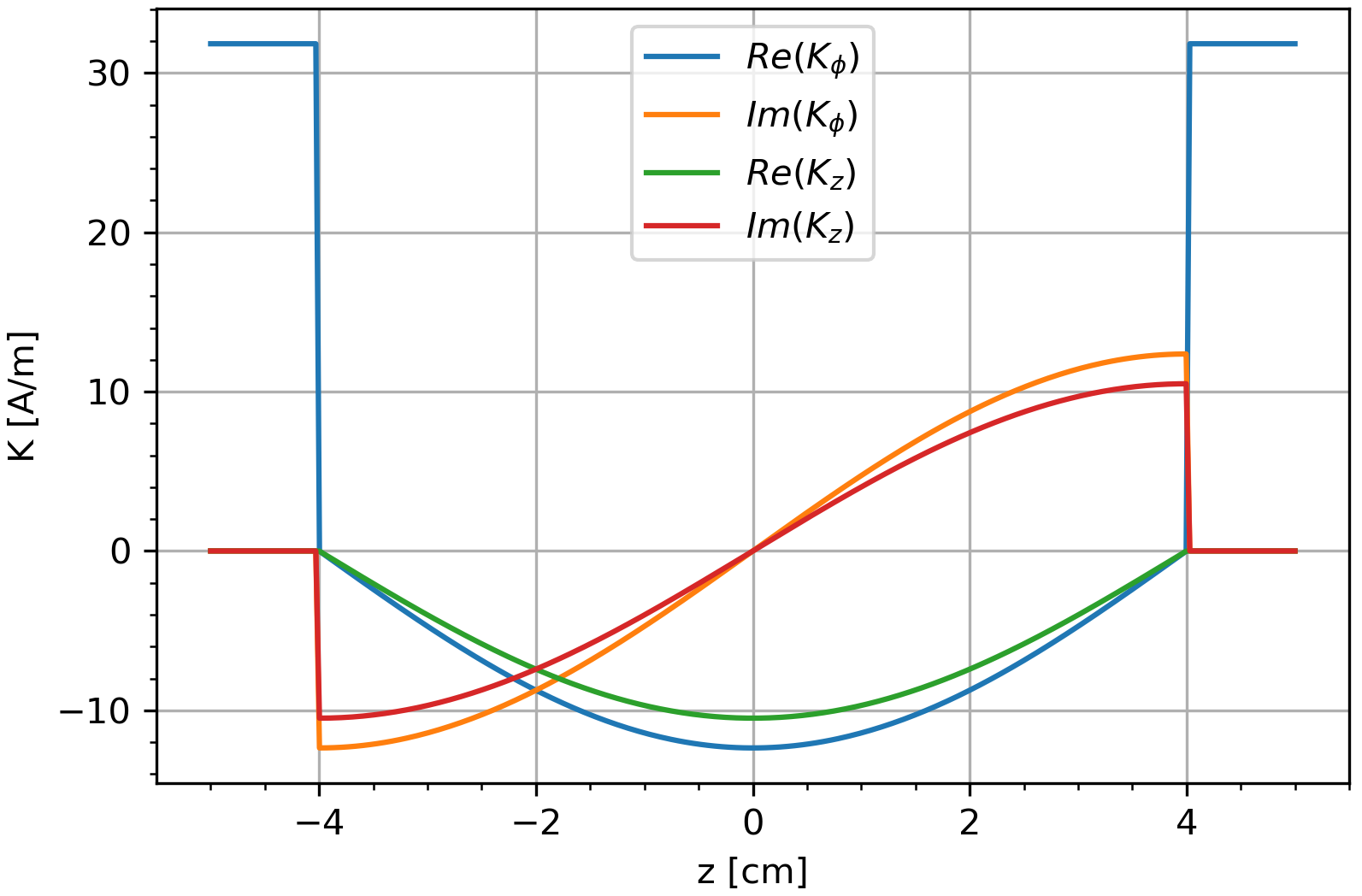}
			\caption{\label{fig:mzSpaceAntennaExamples}Real and imaginary parts of the $K_{\phi}$ and $K_{z}$ current densities for the right-handed antenna on the lower right in \cref{fig:MAPCAD} for azimuthal mode number m=1, calculated with \cref{eq:antCurZ,eq:antCurPhi}.}
\end{figure}

A similar transform was used in \cite{Piotrowicz2018} but there the currents were simplified as delta functions. After a longer calculation, shown in \cref{app:antCurMz}, we find the following surface currents densities in $(m,\,z)$ space

\begin{align}
\label{eq:antCurZ}
\sft{K}_z &= -\frac{I_0 \hel}{\pi R}\ftSinc{m \phi_w} \exp\left(-\frac{i m \pi \hel z }{L_h}\right)\rect{\frac{z}{L_h}}\\
\label{eq:antCurPhi}
\sft{K}_{\phi} &= \gamma \hel \sft{K}_z + \frac{I_0}{m \pi d_t} (-1)^{\frac{m-1}{2}} \left[\rect{\frac{z-z_L}{d_t}}+\rect{\frac{z-z_R}{d_t}}\right],
\end{align}

where we have used the definition $\sinc(x) = \sin(\pi x)/(\pi x)$. Importantly \cref{eq:antCurZ,eq:antCurPhi} apply only to azimuthal modes with odd $m$ and there are no currents in even $m$ modes. An example of these mode currents in $(m, z)$ space is shown in \cref{fig:mzSpaceAntennaExamples} for the $m = 1$ mode.\\

\begin{figure}
\includegraphics[width=\linewidth]{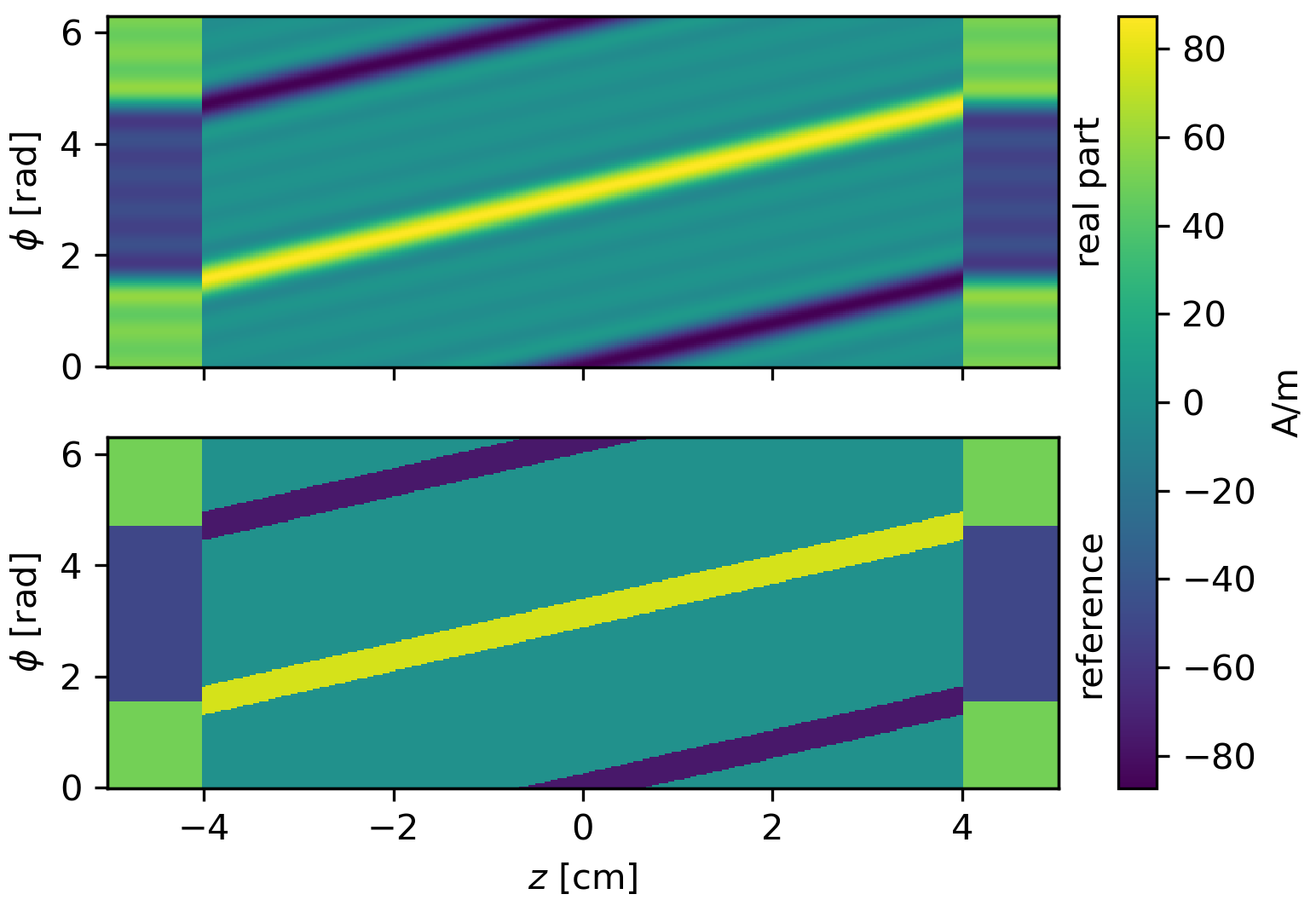}
			\caption{\label{fig:antennaCurrentRec2D}Reconstruction of azimuthal antenna currents $K_{\phi}$ in $(\phi, z)$ space for the ten lowest order azimuthal modes and comparison to the real antenna currents. The imaginary parts of the currents in every $\pm m$ pair cancel each other. The currents have been calculated using \cref{eq:antCurZ,eq:antCurPhi} and summed from $m=-5$ to $m=5$ using the Fourier series in \cref{eq:mphiFtDef2}.}
\end{figure}

\Cref{fig:antennaCurrentRec2D} shows a reconstruction of the azimuthal components of the antenna currents using the first ten modes in the Fourier series. The currents are reconstructed fairly well in magnitude and location using this subset and neglecting higher-order modes. We can therefore expect to model the helicon wave fields accurately by accounting only for these first ten modes.

\subsection{Axial Fourier Transform of Antenna Currents}

We can take the results from \cref{sec:AziFT} and perform a Fourier transform between axial coordinate $z$ and axial wavenumber $k$ to find the antenna power spectrum which will help establish an analytical starting point for antenna optimization later in \cref{sec:Discussion}. To do so we use the definitions

\begin{align}
\dsft{f}(k) &= \frac{1}{2\pi} \int^{\infty}_{-\infty} \sft{f}(z) \ftFac{k z} \,dz\\
\sft{f}(z) &= \int^{\infty}_{-\infty} \dsft{f}(k) \iftFac{k z}\, dk.
\end{align}

This transformation, detailed in \cref{app:antCurMK}, yields the following results for the current densities in $(m, k)$ space:

\begin{align}
\label{eq:Kzhmk}
\dsft{K}_z &= - \frac{I_0 L_h \hel}{2 \pi^2 R} \ftSinc{m \phi_w} \sinc\left[\frac{1}{2} \left(\frac{k L_h}{\pi} + \hel m\right)\right]\\
\label{eq:Kphihmk}
\dsft{K}_\phi &= \gamma \hel \dsft{K}_z +\frac{I_0}{m \pi^2} (-1)^{\frac{m-1}{2}} \cos\left(k z_R \right) \ftSinc{k d_t},
\end{align}

Importantly, \cref{eq:Kzhmk} contains the power spectrum peak location in $k$-space and the launch direction for any azimuthal mode. At the $\dsft{K}_z$ peak we have

\begin{align}
    \label{eq:peakLoc}
    1 &= \sinc\left[\frac{1}{2} \left(\frac{k_{peak} L_h}{\pi} + \hel m\right)\right]\\
    \label{eq:peakLoc2}
    \Rightarrow k_{peak} &= -\frac{\psi m \pi}{L_h}.
\end{align}

Positive and negative $m$ modes are therefore launched in opposite directions and these directions are tied to the antenna helicity $\psi$. This results in the launch directions previously indicated in \cref{fig:MAPCAD}.

\subsection{2D-Axisymmetric Finite Element Simulations}

\label{ssec:fullwavemodel}

To compute the helicon and TG wavefields we use the Comsol finite-element framework\cite{ComsolIntro}. Calculation of the wave fields is achieved in six main steps:

\begin{enumerate}
\item Implementation of the experiment's geometry and assumed or measured plasma density, temperature, and neutral profiles in a 2D-axisymmetric setup.
\item Calculation of the background magnetic field from the coil assembly.
\item Calculation of the dielectric tensor modified by collisions.
\item Solving for the RF wave fields in the frequency domain for discrete azimuthal modes.
\item Combination of azimuthal modes into the 3D solution.
\item Calculation of plasma impedance and scaling of wavefields and power deposition by RF input power.
\end{enumerate}

In the following, we will discuss the process in detail.

\paragraph{Simulation Geometry}

\begin{figure*}
\includegraphics[width=0.9\linewidth]{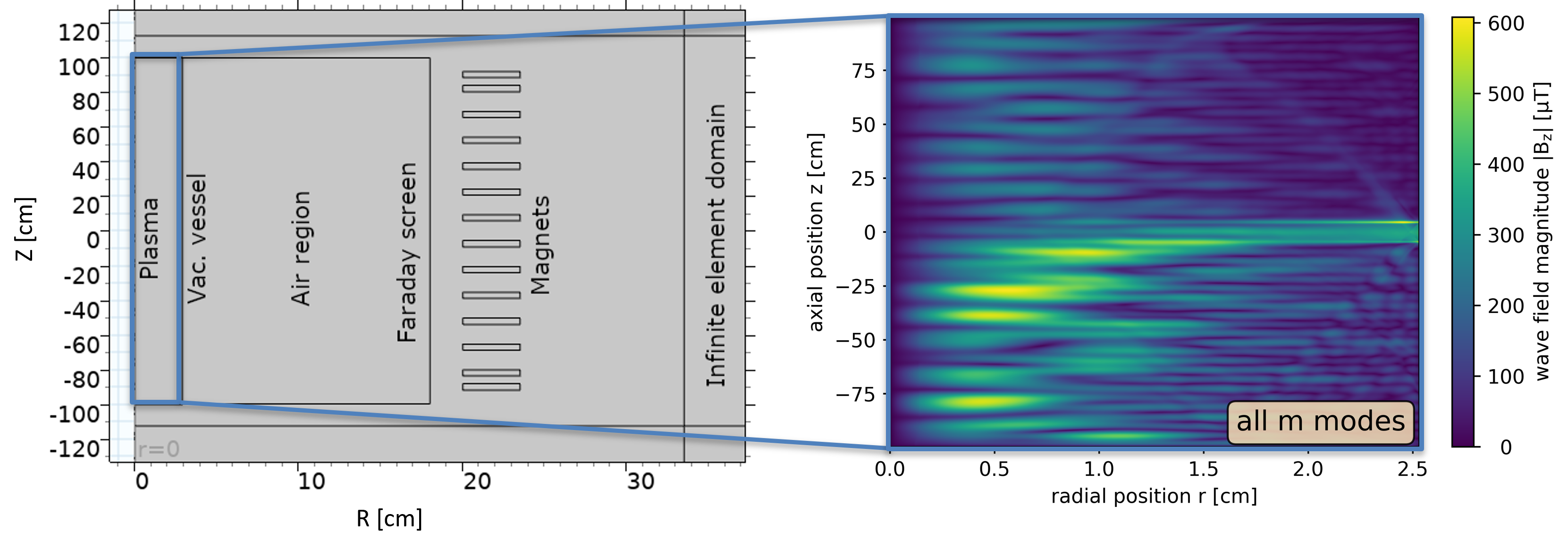}
\caption{\label{fig:ComsolGeometry}Representation of the MAP geometry in Comsol and example of the wavefield solution, in this case for the RMS value of the $B_z$ field summed over all azimuthal modes. The individual azimuthal modes for this simulation are shown later in \cref{fig:ModeExamples}.}
\end{figure*}

\Cref{fig:ComsolGeometry} shows the representation of MAP's geometry in the Comsol framework. Innermost is the vacuum vessel containing the plasma. The walls of the vacuum vessel are modeled as Borosilicate at the radial boundary and perfect conductors at the end caps. The vacuum vessel is surrounded by a Faraday screen, modeled as a perfect conductor. The 14 coils creating the background field are located outside the Faraday screen. The entire arrangement is surrounded by an infinite element domain which is necessary to create a reliable solution during the background magnetic field calculation.\\

\paragraph{Magnetic Field Calculation}

The magnetic background field is calculated from the coil currents using Comsol's ACDC module. At the typical MAP operating point, the field is almost completely homogeneous at 50 mT in the axial direction inside the plasma region and rises slightly above 54 mT at the ends\cite{Granetzny_2024}.\\

\paragraph{Dielectric Tensor}

Given the excitation frequency, electron density, and magnetic field vector the Comsol plasma module is used to calculate the cold plasma dielectric tensor\cite{Stix1992} for each point inside the vacuum vessel assuming a singly ionized argon plasma. Electron-ion and electron-neutral collision frequencies are calculated from temperature-dependent cross-section data provided by Comsol\cite{Comsol-ECR}. Collisions are included in the dielectric tensor by replacing $m_e$ with $m_e\left(1+\frac{i\nu}{\w}\right)$\cite{Chen1997}, where $\nu$ is the combined collision frequency.\\

\paragraph{RF Mesh}

An important characteristic of the helicon and TG waves is the large difference between radial and axial wavelengths as demonstrated earlier in \cref{fig:kBetaPlot}. For Comsol RF simulations it is generally recommended to have about 10 mesh elements per wavelength. We therefore reduce computational cost by creating elements with very high aspect ratio thereby exploiting the order of magnitude difference between radial and axial wavelengths as demonstrated in \cref{fig:kBetaPlot}.\\

\paragraph{RF Solver}

Solving the Maxwell equations is done in the frequency domain in cylindrical coordinates under the assumption that any wavefield $F(r, \phi, z, t)$ is of the form

\begin{align}
F(r, \phi, z, t) &= f(r, z)e^{i\left(m\phi - \w t\right)},
\end{align}

with $\w$ being the wave's angular frequency and $m$ being the azimuthal mode number discussed earlier. Azimuthal and time derivatives then simplify as

\begin{align}
\frac{\partial}{\partial \phi} &\rightarrow  i m\\ 
\frac{\partial}{\partial t} &\rightarrow - i \w.
\end{align}

The wave fields are driven by the complex-valued antenna surface currents which are implemented according to \cref{eq:antCurZ,eq:antCurPhi} for a total current of $I_0 = 1 \U{A}$. For each azimuthal mode number $m$ this current density is applied at the outer vacuum vessel boundary at the antenna's axial position as exemplified earlier in \cref{fig:mzSpaceAntennaExamples}. While the simulations shown here are limited to a single antenna at $z = 0$, additional antennas with different positions and phases can be added. To adjust any antenna's position we replace $z$ in \cref{eq:antCurZ,eq:antCurPhi} with $z - z_0$, with $z_0$ being the new position. To introduce a phase shift $\phi_0$, we multiply \cref{eq:antCurZ,eq:antCurPhi} by a factor $e^{i\phi_0}$. The wavefields for all computed azimuthal modes are then exported for post-processing.\\

\paragraph{Reconstruction of 3D Wavefields}

The wave fields of the 3D antenna can be reconstructed from the individual azimuthal modes in the same way that the 3D antenna current is reconstructed according to \cref{eq:mphiFtDef2}. For any wave field quantity $F(r, \phi, z, t)$ we have

\begin{align}
\label{eq:fieldRec}
F(r, \phi, z, t) &= \ftFac{\w t}\sum_{m = -\infty}^{\infty} F_m(r, z) \iftFac{m \phi}
\end{align}

To visualize the wavefields it is beneficial to calculate the root-mean-square value of \cref{eq:fieldRec}, which is found as

\begin{align}
\label{eq:modeCombination}
\avg{F}_{rms} &= \sqrt{\sum_{m = -\infty}^{\infty} |F_m|^2}.
\end{align}

\paragraph{Calculation of Power Deposition}

An important result of the computational model is the capability to predict the resistive power deposition inside the plasma as well as the plasma resistance and reactance.\\

The cycle-averaged local power deposition density $P$ in the plasma is simply

\begin{align}
\label{eq:powerDepDensity}
P &= \conj{\vE}\cdot \vj\\
\label{eq:powerDepDensityByMode}
&= \sum_m \conj{\vE_m}\cdot \vj_m,
\end{align}

The total plasma impedance $Z$ can be calculated directly from the total volume-integrated power deposition in the plasma $P_{tot}$ as

\begin{align}
\label{eq:impedanceCalc}
P_{tot} &= \frac{1}{2}I_0^2 Z,
\end{align}

where $I_0$ is the amplitude of the antenna current.\\

\paragraph{Scaling of Wavefields with Antenna Current}

In practice, the power level in \cref{eq:impedanceCalc} will be set at the RF generator and we can assume the power to be almost completely absorbed in the plasma due to an impedance-matching network. The plasma impedance $Z$ is a property of the plasma itself and not dependent on the wavefields in it. Therefore, the current amplitude $I_0$ will adjust to the plasma conditions and input power levels. The matching network will compensate for the plasma's reactance, so only the resistive (i.e. real) part of the power needs to be evaluated. Since the simulations are performed for 1 A antenna current the real antenna current for a given input power level can be calculated as\cite{Curreli2011}

\begin{align}
I_{real} &= 1\U{A}\times\sqrt{\frac{P_{in}}{Re(P_{sim})}}.
\end{align}

Since the Maxwell equations and the cold plasma model are linear, all fields scale linearly with the antenna current.

\section{Results}
\label{sec:Results}

\begin{figure}
\includegraphics[width=\linewidth]{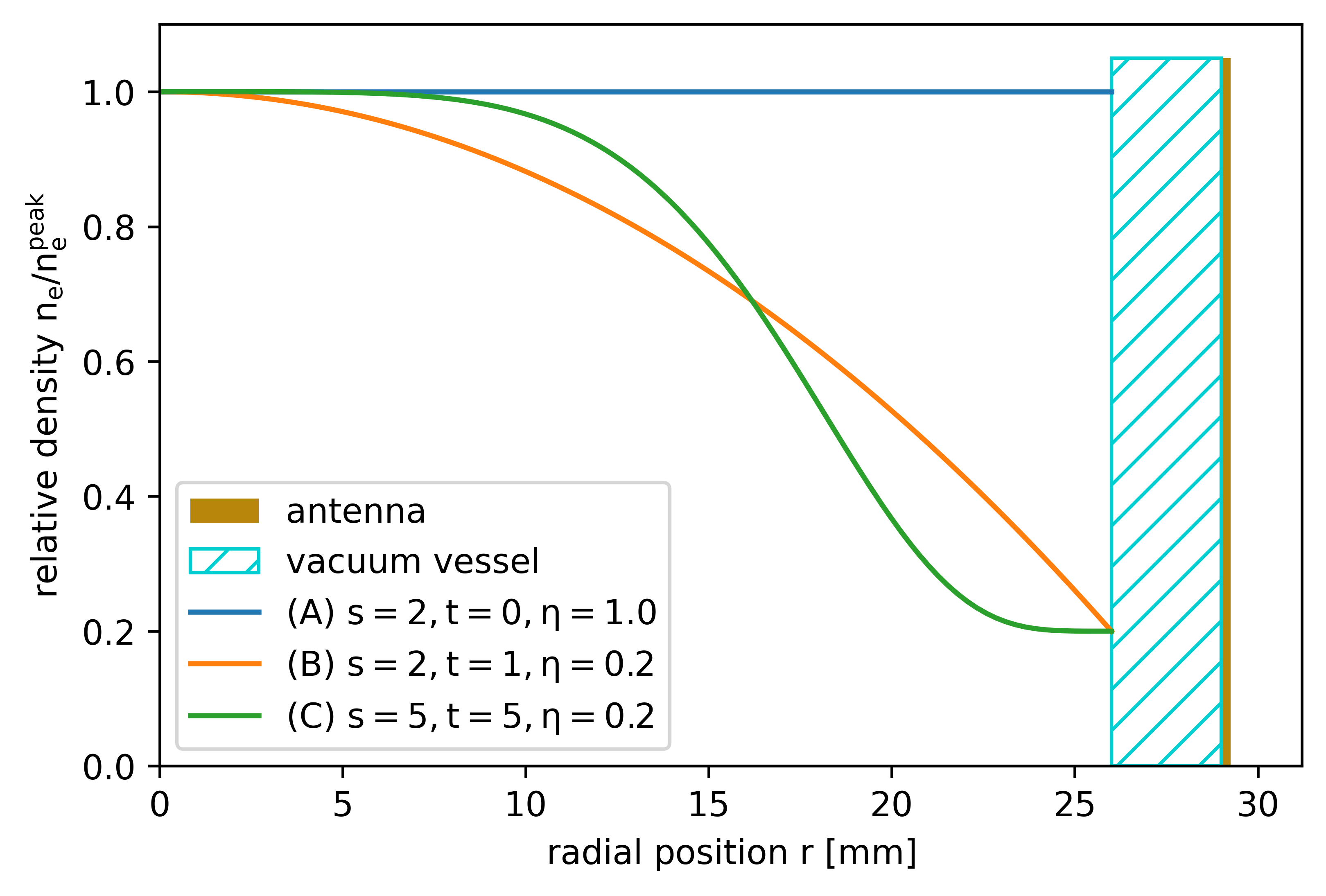}
\caption{\label{fig:shapes}Radial plasma density profiles used throughout this work, corresponding to different parameter choices for $s$, $t$, and $\eta$ in \cref{eq:shape}.}
\end{figure}

The subsequent results were derived for axially uniform plasmas with different radial density profiles. These profiles are all of the form

\begin{align}
\label{eq:shape}
n_e &= \pbr{(1 - \eta)\left[1-\left(\frac{r}{a}\right)^s\right]^t + \eta}n_e^{peak},
\end{align}

where $s$ and $t$ define the profile shape and $\eta$ is the edge density as a fraction of the peak density. The profiles used throughout the rest of this work are shown in \cref{fig:shapes}. (A) is a completely uniform plasma, (B) is a parabolic profile and (C) is a flat-top profile. The vacuum vessel in \cref{fig:shapes} is the one used at the MAP experiment with an inner radius of $a = 26\U{mm}$ and a wall thickness of $3\U{mm}$. The outermost layer at $r = 29\U{mm}$ in \cref{fig:shapes} is the location at which the antenna currents are applied according to \cref{eq:antCurZ,eq:antCurPhi}.

\subsection{Significant Modes}

\begin{figure}
\includegraphics[width=\linewidth]{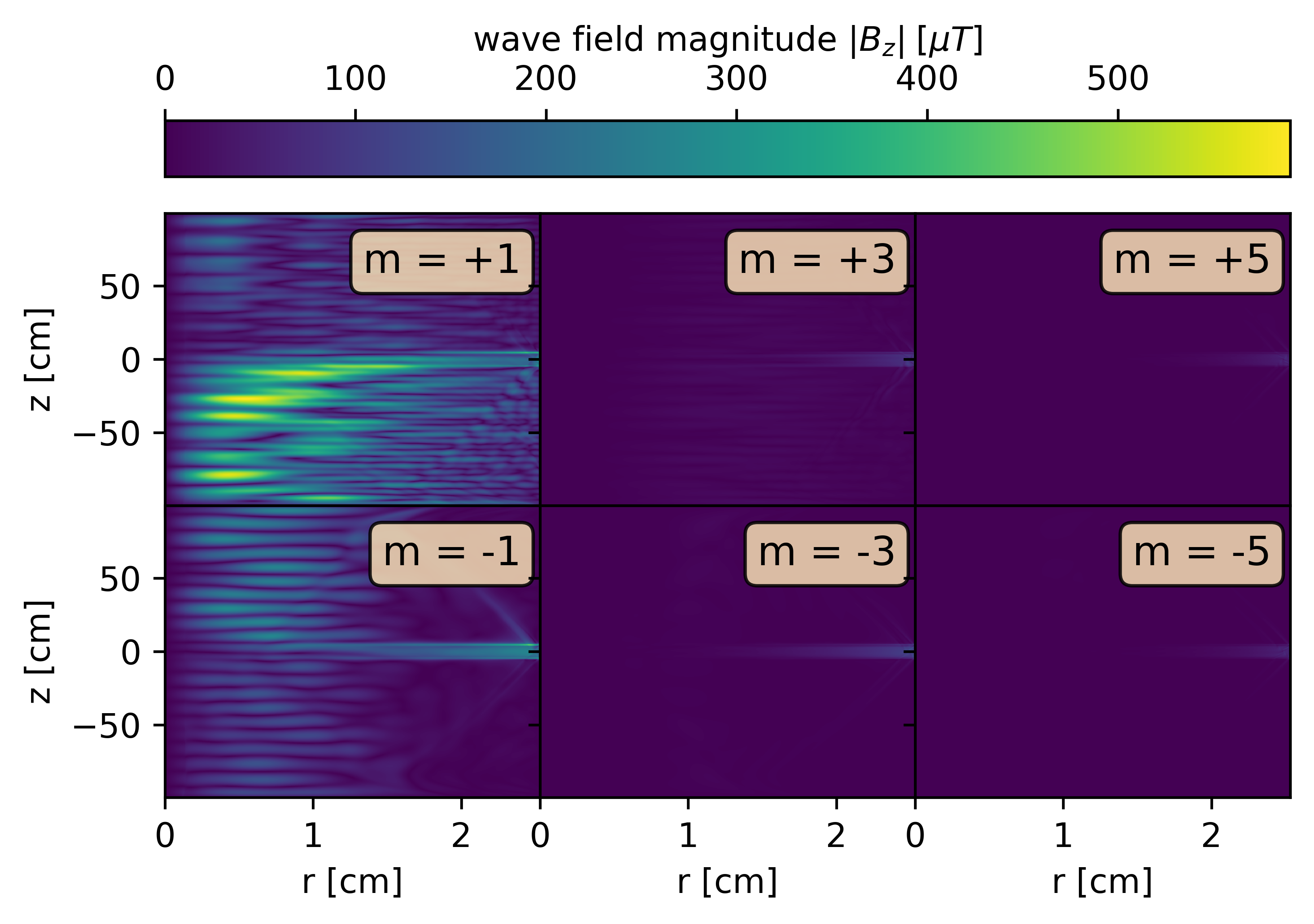}
\caption{\label{fig:ModeExamples}$|B_z|$ wave fields for first six azimuthal modes in an axially uniform plasma with radial density flat top density profiles of type (C) in \cref{fig:shapes}. The first two modes, $m=\pm 1$, show significant contribution to the overall wavefield while higher order modes up to $m \pm 5$ yield slight corrections.}
\end{figure}

The Fourier decomposition strategy described above allows us to solve for an arbitrarily high number of azimuthal modes. However, most of these will be strongly damped and make only a small contribution to the overall wavefields and power deposition. To demonstrate this effect we show the $B_z$ wavefield components for the six highest order modes in \cref{fig:ModeExamples}. These simulations were performed for plasmas with a uniform temperature of $3\U{eV}$ and a neutral gas pressure of $\Eb{-3}\U{mbar}$. The plasma density was assumed to be axially uniform with a radial flat top profile with finite edge density of form (C) in \cref{fig:shapes}.\\

We find that only the first two azimuthal modes, $m=\pm 1$, have significant contributions to the overall wavefield, with the next four modes, $m=\pm 3$ and $m=\pm 5$, providing small corrections. Higher order modes such as the $m=\pm 7$ mode are even weaker and can be safely neglected. We therefore only need to perform 2D-axisymmetric simulations of the first four or six modes to accurately reconstruct the full 3D wavefields. The 3D reconstruction for this simulation, using \cref{eq:modeCombination}, has been shown previously on the right in \cref{fig:ComsolGeometry}.

\subsection{Comparison to Analytical Dispersion Relation}

\begin{figure}
\includegraphics[width=\linewidth]{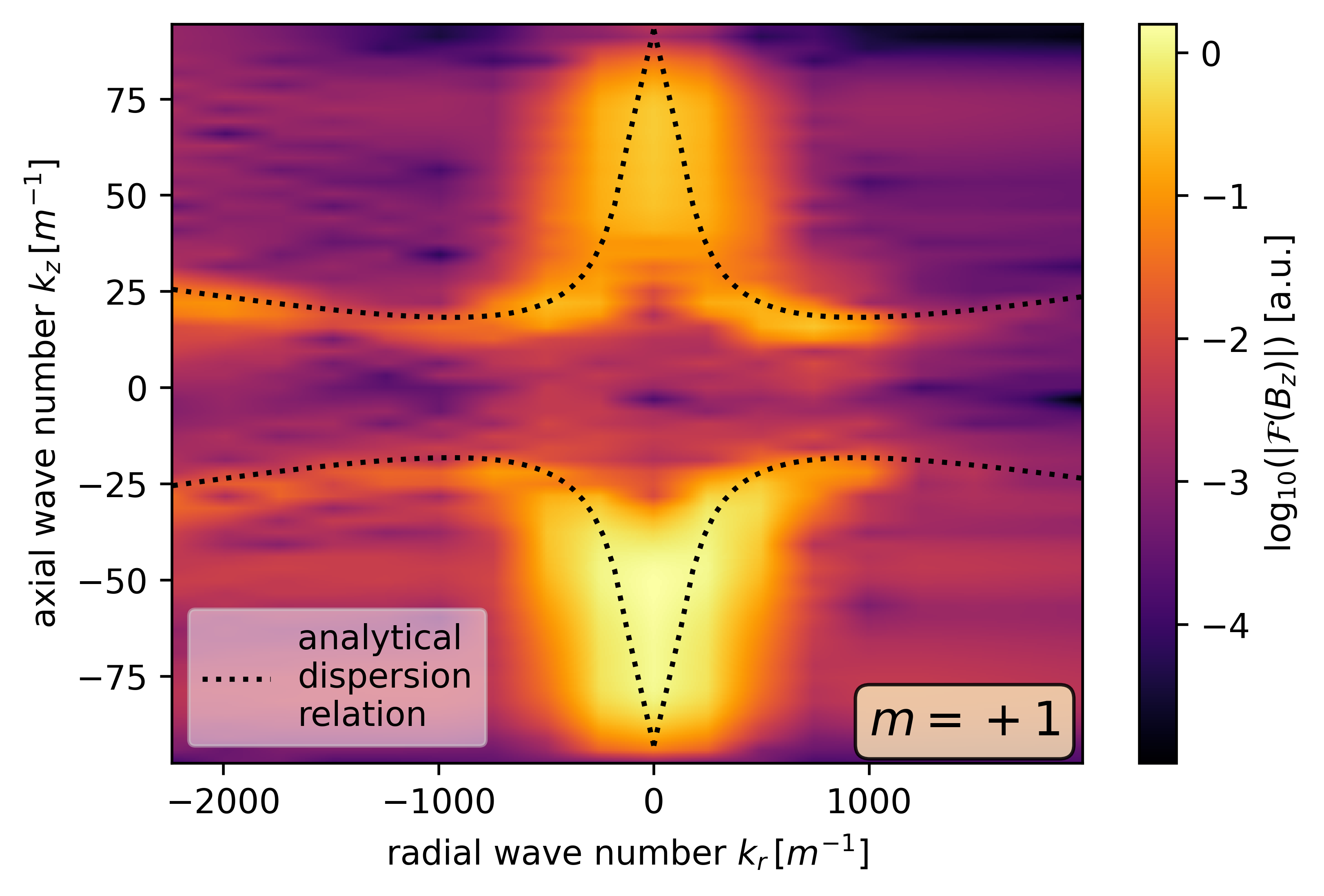}
\caption{\label{fig:uniSpecPara}Comparison of the magnitude of the Fourier transform of the $B_z$ field component in the $m=+1$ mode to the analytical dispersion relation in a uniform plasma with $n_e = 2.5\E{19}\dens$, $B_0 = 50 \U{mT}$, $f=13.56 \U{MHz}$ driven by a 10 cm long antenna with 1 cm wide straps. The magnitude is plotted on a logarithmic color scale in arbitrary units.}
\end{figure}

To validate the model, we can run it in a completely uniform plasma. A Fourier analysis of the wavefields should then recover the dispersion relation of \cref{eq:heliconDispRel}. We performed this test for a plasma with a uniform density of $2.5\E{19}\U{\dens}$ in a homogeneous 50 mT field with a 10 cm long right-handed antenna. The two-dimensional Fourier transform for the $B_z$ wavefields of the $m=1$ mode is shown in \cref{fig:uniSpecPara} on a logarithmic scale. The black curves represent the analytical dispersion relation according to \cref{eq:heliconDispRel}. Unlike the analytical dispersion relation, the numerical Fourier spectrum has peaks of finite width due to the bounded nature of the simulated plasma and the finite resolution provided by the mesh elements. Forward (positive $k_z$) and backward (negative $k_z$) as well as inward (negative $k_r$ and outward (positive $k_r$) propagating waves are visible. The computational results are in agreement with the analytical dispersion relation. We further find that the TG branch is much weaker than the helicon branch as we would expect due to the strong damping of the former.\\

\subsection{Antenna Optimization}

\begin{figure}
\includegraphics[width=\linewidth]{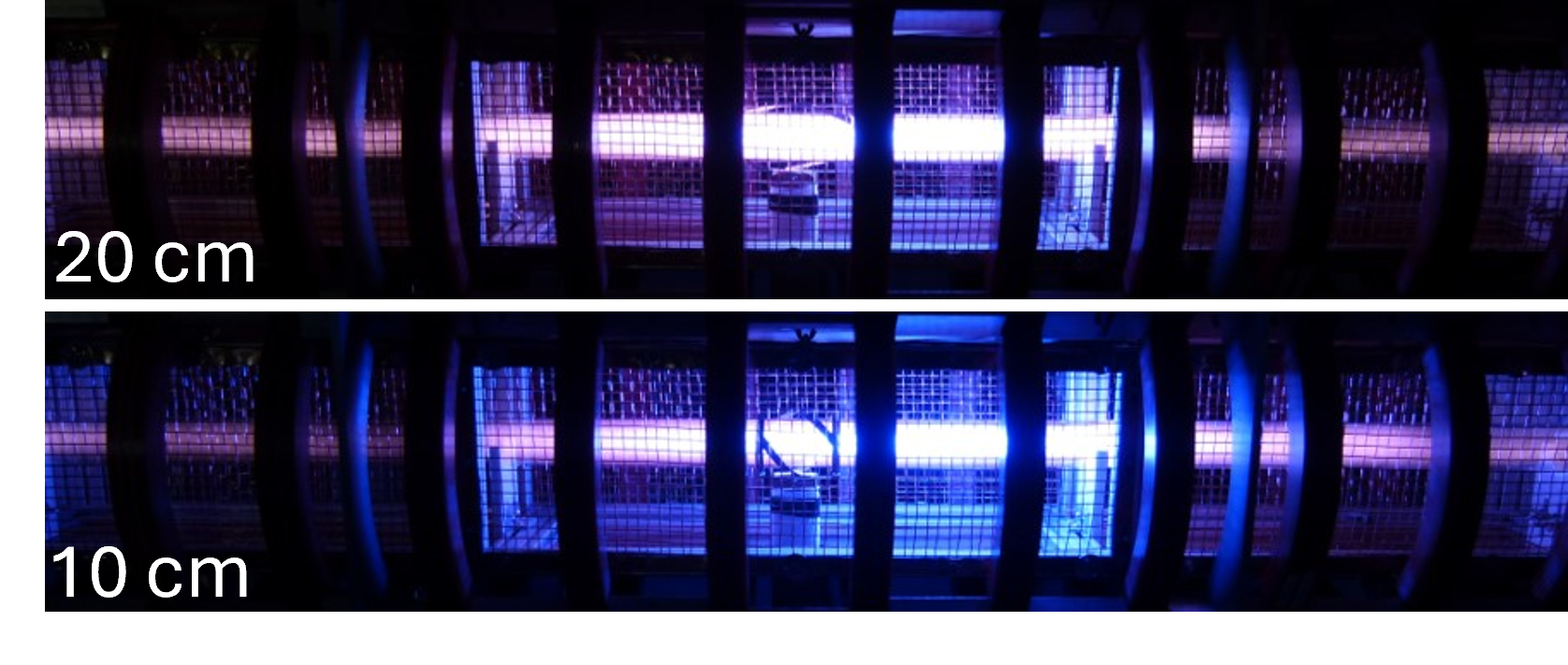}
\caption{\label{fig:lengthExp}Comparison of MAP plasmas using antennas of two different lengths with otherwise identical operating conditions at $50 \U{mT}$ and argon fill pressure of 1 Pa. The top panel shows operation with a 20 cm antenna. The discharge is mainly purple and axially symmetric, indicating mainly neutral gas rather than ion excitation and an inductively coupled discharge. The bottom panel shows operation with a 10 cm antenna. The plasma has a strongly directional blue core, indicating helicon operation and good power coupling.}
\end{figure}

It is experimentally known\cite{Granetzny_2023} that helicon plasmas are strongly directional. This effect is clearly visible as a strong asymmetry in light emission and density profiles around the antenna location, as shown for example in \cref{fig:lengthExp}. The top panel shows operation with a 20 cm antenna. The discharge is mainly purple and axially symmetric, indicating argon neutral gas excitation and an inductively coupled plasma. In contrast, the bottom panel shows operation with a 10 cm antenna. The plasma has a strongly directional blue core, indicating helicon operation with strong excitation of ArII ions and good power coupling. It was shown previously that this directionality is fundamental to laboratory helicons and arises from the interaction between the radial electric wavefields with the radial density gradient and background magnetic field.\cite{Granetzny_2023} In \cref{fig:lengthExp} the magnetic field points to the left (towards negative $z$) such that the waves propagate predominantly to the right unlike in computational results shown here which all use a field pointing to the right.\\

\begin{figure}
\includegraphics[width=\linewidth]{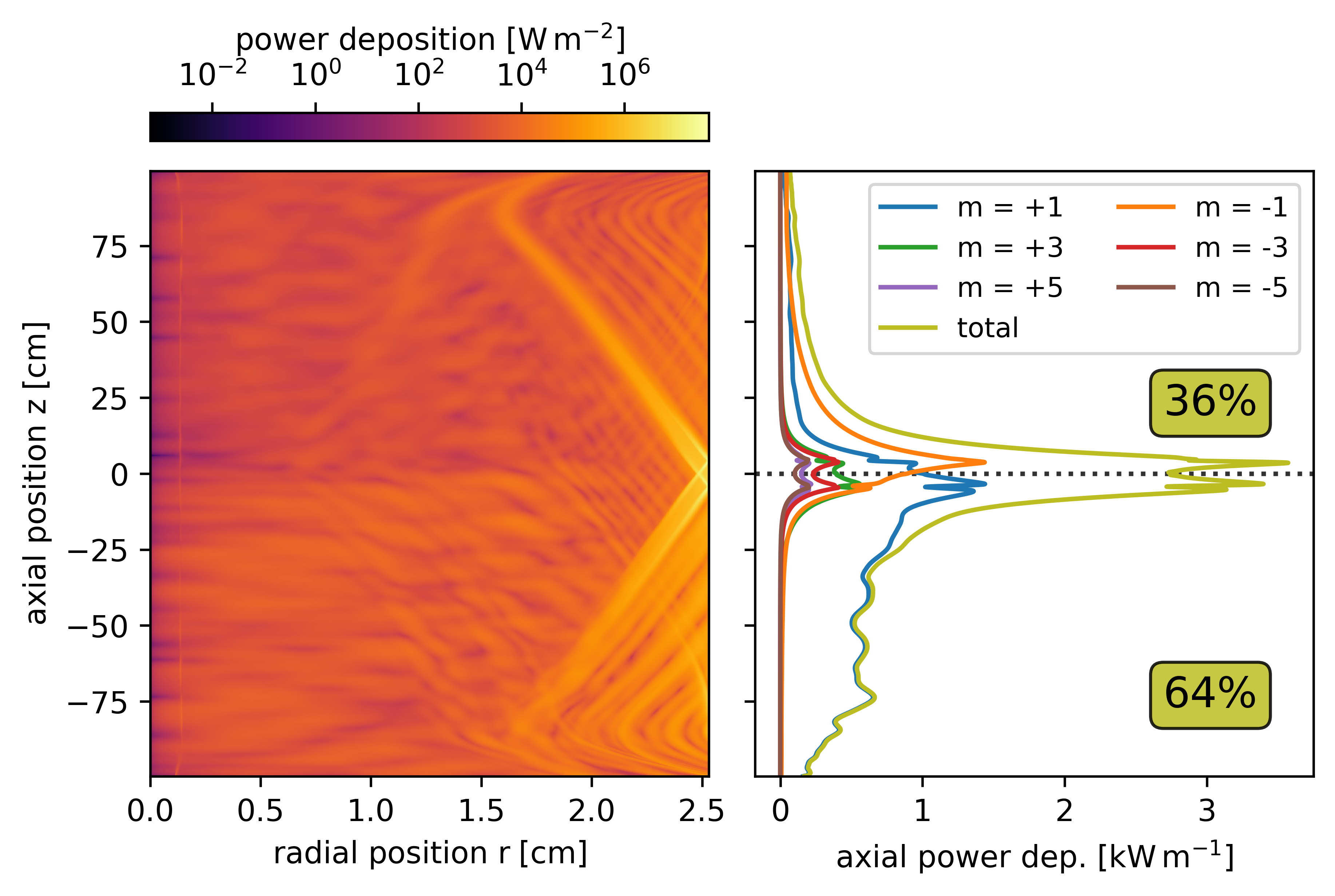}
\caption{\label{fig:powerDepExample}Power deposition by the wavefields shown previously in \cref{fig:ComsolGeometry,fig:ModeExamples}. Left: Power deposition is mostly located at the edge due to strong TG mode damping. Right: Radial integration of the power deposition shows that 64\% of the power is towards negative $z$ values, corresponding to waves propagating predominantly in that direction. Further, the majority of the power deposition is due to the $m\pm1$ modes as expected from \cref{fig:ModeExamples}.}
\end{figure}

This axial power deposition asymmetry corresponds to asymmetric wavefields, as shown previously on the right in \cref{fig:ComsolGeometry}, where the antenna is located at $z = 0$ and most of the wavefields are in the negative $z$ region. This imbalance leads to asymmetric power deposition around the antenna as shown in \cref{fig:powerDepExample} where 64\% of power deposition occurs in the negative $z$  direction. In general, a stronger asymmetry indicates a better coupling of RF power from the antenna into the plasma.\\

\begin{figure*}
\includegraphics[width=0.8\linewidth]{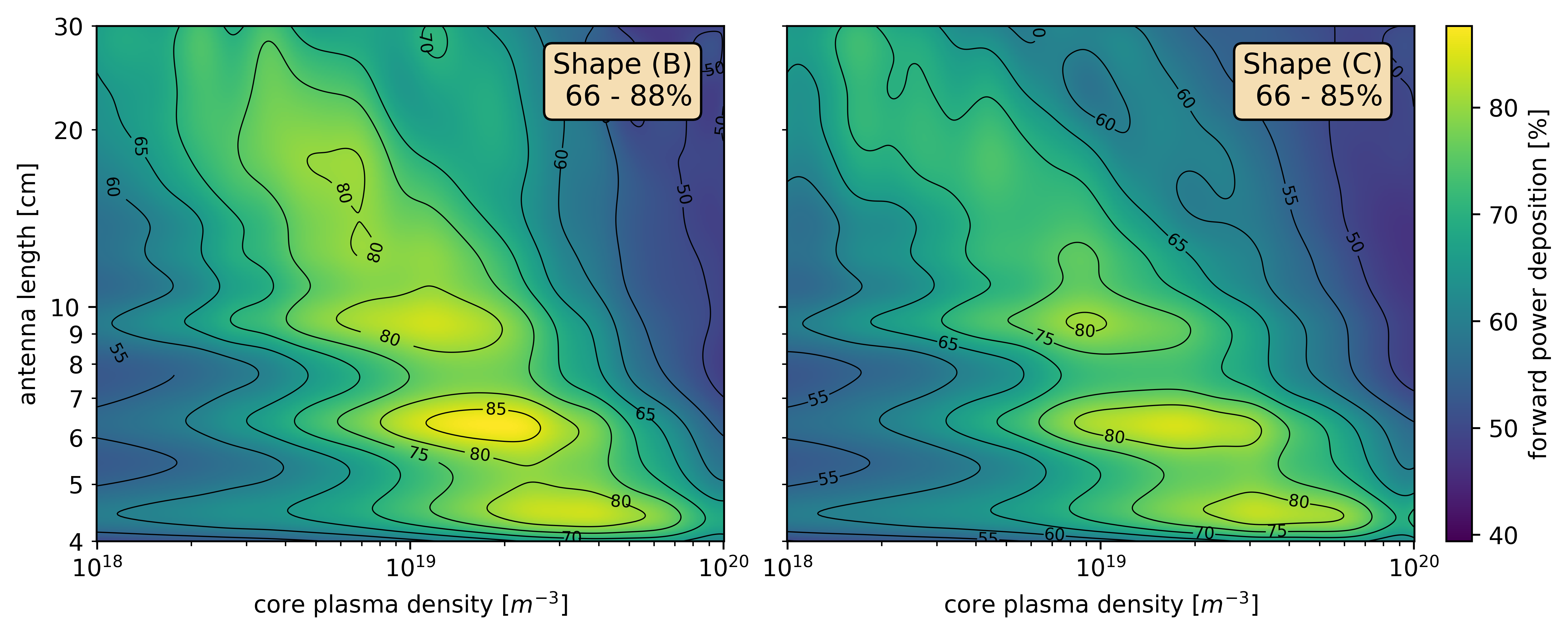}
\caption{\label{fig:powerRatio}Power deposition asymmetry for axially uniform plasmas with core densities from $\ebd{18}$ to $\ebd{20}$ and antenna lengths from 4 to 30 cm. Radial densities profiles correspond to cases (B) and (C) in \cref{fig:lengthExp}.} 
\end{figure*}

Based on these experimental findings, we expect antennas of different lengths to yield significant differences in power coupling and overall helicon plasma performance. To investigate this issue, we have conducted simulations for different combinations of antenna length from 4 to 30 cm and core plasma densities - $n_e^{peak}$ in \cref{eq:shape} - from $\ebd{18}$ to $\ebd{20}$. These plasmas were axially uniform and had radial density profiles of the form (B) or (C) in \cref{fig:shapes}. Electron temperature and neutral pressure were set uniformly to $3\U{eV}$ and $\Eb{-3}\U{mbar}$, respectively. The result is shown on the left in \cref{fig:powerRatio} for 400 different combinations of antenna length and core densities for radial density profiles of type (B). The contour plot shows the power deposition asymmetry as a proxy for power coupling efficiency with higher values indicating better coupling. A value of 50\% means that the power deposition is symmetric, whereas a value of 100\% means that all power is deposited to the preferred side of the antenna. We find that for any given core density there exists an optimal antenna that will couple 66-88\% of power into the preferred side of the plasma and that the ideal antenna length is strongly dependent on the core density. The optima trace out a ridge, with higher core densities requiring smaller antenna lengths.\\

We conducted the same study with the profile shape (C) in \cref{fig:shapes}. The results are shown on the right in \cref{fig:powerRatio}. The differences in the power coupling between profiles (B) and (C) landscape are small and mostly show as a narrowing of the diagonal ridge, thus indicating that the exact radial profile shape has only a small impact on power coupling efficiency. The optimized antennas result in power coupling efficiency ranges from 66-85\%, almost the same as for profile shape (B).\\

\section{Discussion}
\label{sec:Discussion}

% Tracing the optima in \cref{fig:powerRatio}, we can find the ideal antenna length for a given core density, as shown in \cref{fig:ratioComp}. For any given density, the optimal antenna length is the same or nearly the same for both radial profiles. The relationship between core density and antenna length length fits a simple exponential scaling law of the form

% \begin{align}
% \label{eq:optFitNe}
%     n_{e}^{peak}[\ebd{19}] &= 0.28 + 32\times \exp\br{-0.44 L [\U{cm}]}\\
% \label{eq:optFitL}
%     \Leftrightarrow L[\U{cm}] & = 2.3 \times \ln\br{\frac{32}{n_{e}^{peak}[\ebd{19}] - 0.28}}.
% \end{align}

\begin{figure*}
\includegraphics[width=1.0\linewidth]{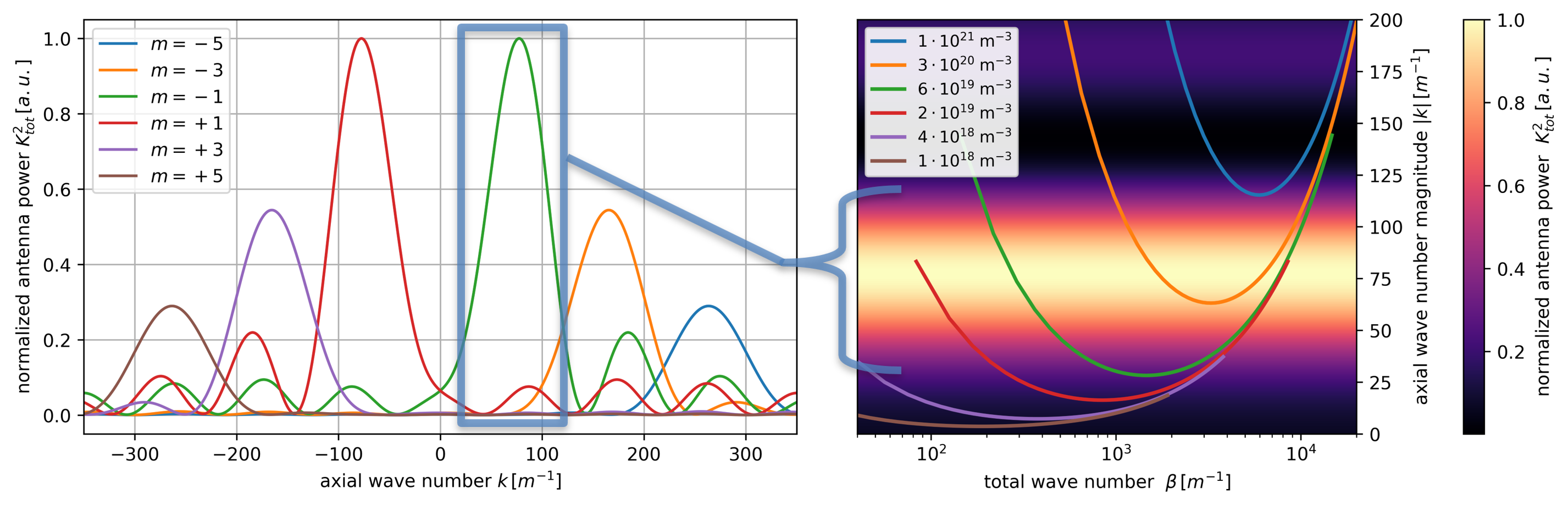}
\caption{\label{fig:anaOpt}Overlap between the antenna power spectrum of an 8 cm long antenna (left) and the helicon dispersion relation at different densities ranging from $\ebd{18}$ to $\ebd{21}$ (right). Power coupling can be optimized by designing the antenna for a large overlap of wavenumbers excited by the dominant $m = \pm 1$ modes with the dispersion relation at the densities of interest.}
\end{figure*}

Tracing the optima in \cref{fig:powerRatio}, we find that the ideal antenna length for a given core density is the same or nearly the same for both radial profiles. The existence of such an optimum can be understood by comparing the antenna power spectrum with the helicon dispersion relation. We can calculate the power spectrum $K_{pow}$ in wavenumber space as 

\begin{align}
    \dsft{K}_{pow} &= \sqrt{\left|\dsft{K}_z \right|^2 + \left|\dsft{K}_\phi \right|^2},
\end{align}

using equations \cref{eq:Kphihmk,eq:Kzhmk}. An example of such a power spectrum is given on the left in \cref{fig:anaOpt} which shows the different azimuthal modes in dependence on the axial wavenumber in an 8 cm long antenna. Crucially, the helicon-TG dispersion relation in \cref{eq:heliconDispRel} defines which axial wavenumbers are permissible in a uniform plasma at a certain density, field strength, and frequency. The right side in \cref{fig:anaOpt} shows this dispersion relation for plasma densities ranging from $\ebd{18}$ to $\ebd{21}$. We know experimentally and from our simulation results in \cref{fig:ModeExamples,fig:powerDepExample} that only $m \pm 1$ modes contribute significantly to wavefields and power deposition in helicon plasmas. We can therefore focus on the $m = 1$ peak in \cref{fig:anaOpt} and compare the excited wave numbers with those allowed by the dispersion relation at different densities relevant to our plasma, as shown on the right in \cref{fig:anaOpt}.\\

In this example, the antenna power spectrum has a partial overlap with the dispersion relation at $2\ed{19}$ and $3\ed{20}$, shown by the red and orange curves, respectively. At $6\ed{19}$, the full power spectrum for the $m=1$ mode intersects the dispersion relation. In contrast, the overlap is negligible for the lowest densities in the $\ebd{18}$ range, shown by the purple and brown curves. The same is true at $\ebd{21}$. We can therefore expect that such an antenna will optimize power coupling into a mid-$\ebd{19}$ plasma but will lead to a non-optimized discharge at lower or higher densities.\\

For a given target density, the optimal antenna length can be estimated as follows. The dispersion relation in \cref{eq:heliconDispRel} allows axial wave numbers up to $k = \beta$ in which case the wave propagates fully in the axial direction. Inserting this condition back into \cref{eq:heliconDispRel} we get

\begin{align}
\label{eq:kmax}
    k_{max} &= \frac{k_w}{\sqrt{1 - \delta}}.
\end{align}

On the other hand the dispersion relation is parabolic with a minimum at 

\begin{align}
\label{eq:kmin}
    k_{min} &= 2 k_w \sqrt{\delta}.
\end{align}
    
A good first estimate for the ideal antenna length is then to engineer a dominant power spectrum peak around the middle of the axial wavenumber range allowed by the dispersion relation. We can set this location by means of a parameter $\alpha$, anywhere between $k_{min}$ ($\alpha = 0$) and $k_{max}$ ($\alpha = 1$), with ($\alpha = 0.5$) resulting in a peak right at the midpoint. Referring back to \cref{eq:peakLoc2,eq:LhDef} we have

\begin{align}
\label{eq:optIdea}
    \left| -\frac{\psi \pi m}{L - 2d_t}\right| &= k_{min} + \alpha\br{k_{max} - k_{min}},
\end{align}

where we have for now ignored modifications of the power spectrum due to the transverse strap currents in \cref{eq:Kphihmk}.\\

We know that the $m = \pm 1$ modes are the only significant contributors to the overall power deposition. At the same time, $\psi = \pm 1$ changes the plasma direction but does not change the ideal antenna length. Using these constraints and rearranging \cref{eq:optIdea,eq:kmin,eq:kmax} we find for the ideal antenna length $L_{ideal}$

\begin{align}
\label{eq:Lideal}
    L_{ideal} &= \frac{\pi}{k_w\br{2\sqrt{\delta}\br{1-\alpha} + \frac{\alpha}{\sqrt{1-\delta}}}} + 2d_t,
\end{align}

with $k_w = \sqrt{2\pi f n_e \mu_0 e/B}$ and 
$\delta = 2\pi f m_e/e B$ as defined previously in \cref{eq:kw,eq:delta}, where we have ignored collisions for this analysis.\\

\begin{figure}
\includegraphics[width=\linewidth]{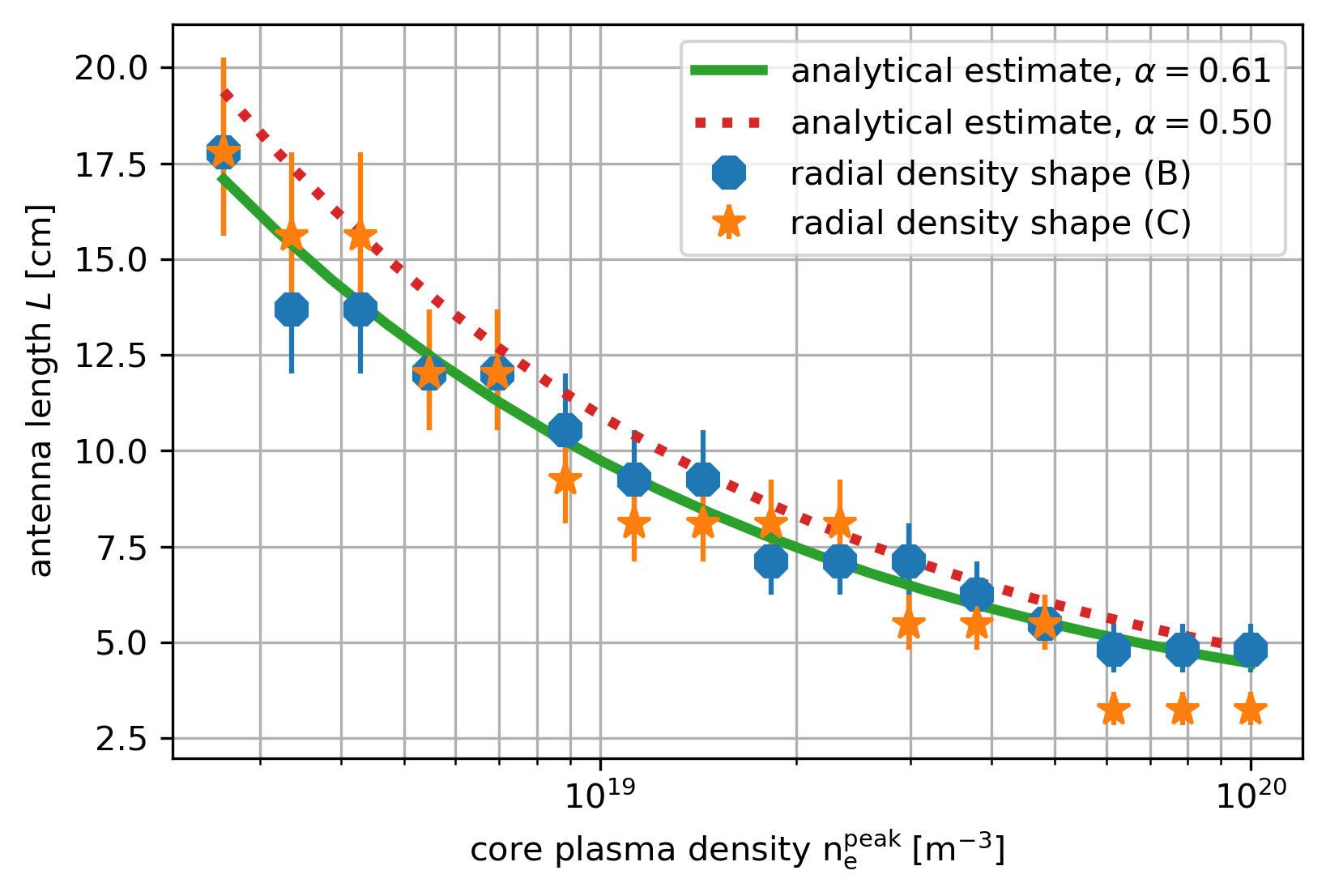}
\caption{\label{fig:ratioComp}Comparison of the ideal antenna length for profiles with different core densities for radial profile shapes (B) and (C) in \cref{fig:shapes}.  The optimal antenna length is the same or nearly the same in both configurations and can be predicted from \cref{eq:Lideal} by engineering a power spectrum peak that has maximal overlap with the dispersion relation at a given density. The red dotted curve shows this relationship for a $\dsft{K}_z$ peak right in the middle between the minima and maxima of the dispersion relation. The solid green curve shows the result of numerical optimization with the spectral peak at slightly higher axial wavenumbers.} 
\end{figure}

Using \cref{eq:Lideal} we can compare our analytical optimization approach to the simulation data in \cref{fig:powerRatio}. \Cref{fig:ratioComp} shows the ideal antenna length for profile shapes (B) and (C). The antenna length for a power spectrum peak at the midpoint of $k_{min}$ and $k_{max}$, i.e. $\alpha = 0.5$, is shown as the red dotted line and is in good agreement with the simulation results. Using the data in \cref{fig:ratioComp} we find a best-fit value of $\alpha = 0.61$. The result is the green curve in \cref{fig:ratioComp} which shows excellent agreement over two orders of magnitude in density and a factor of seven in antenna size.\\

In practice, creating a plasma of a certain density will require fulfilling the particle and power balance\cite{Lechte2002,Zepp_2024}. However, the optimization procedure described here can be used to find the antenna length that minimizes the amount of RF power needed to achieve a certain target density. 

\section{Summary}
\label{sec:Summary}

We have developed a new 3D finite element model for studies of helicon and Trievelpiece-Gould wave propagation and power deposition in linear devices using realistic representations of half-helical antennas. The model uses a Fourier decomposition of the antenna currents to solve the full 3D problem through a small number of 2D-axisymmetric simulations. The resulting wavefield and power deposition patterns are consistent with experimental findings. This new approach leads to a dramatic reduction in the computational resources required compared to a direct 3D simulation. We have used this new capability to optimize helicon power coupling by changing the antenna length in dependence on the target plasma density. The optima are independent of the exact shape of the radial density profile. We are able to reproduce these findings analytically through a comparison between the antenna power spectrum and the helicon dispersion relation. The practical application of this finding is the optimization of power coupling into a helicon plasma by designing a half-helical antenna with a length according to \cref{eq:Lideal}. Such an optimized antenna will minimize the RF power needed to create a helicon plasma with a given target density. We expect these results to be applicable to any linear helicon device with a sufficiently homogenous magnetic field. \\

\ano{The research presented here was funded by the National Science Foundation under grants PHY-1903316 and PHY-2308846 as well as the College of Engineering at UW-Madison.\\}

\ano{
\section*{Author Declarations}
\subsection*{Conflict of Interest}
The authors have no conflicts to disclose.
\subsection*{Author Contributions}
\textbf{Marcel Granetzny: }
\optCredit{Conceptualization}{\ld}
\optCredit{Data curation}{\ld}
\optCredit{Formal analysis}{\ld}
\optCredit{Funding acquisition}{}
\optCredit{Investigation}{\ld}
\optCredit{Methodology}{\ld}
\optCredit{Project administration}{\eq}
\optCredit{Resources}{\su}
\optCredit{Software}{\ld}
\optCredit{Supervision}{}
\optCredit{Validation}{\ld}
\optCredit{Visualization}{\ld}
\optCredit{Writing - original draft}{\ld}
\optCredit[.]{Writing - review \& editing}{\ld}
\textbf{Oliver Schmitz: }
\optCredit{Conceptualization}{\su}
\optCredit{Data curation}{}
\optCredit{Formal analysis}{}
\optCredit{Funding acquisition}{\ld}
\optCredit{Investigation}{}
\optCredit{Methodology}{}
\optCredit{Project administration}{\eq}
\optCredit{Resources}{\ld}
\optCredit{Software}{}
\optCredit{Supervision}{\ld}
\optCredit{Validation}{}
\optCredit{Visualization}{}
\optCredit{Writing - original draft}{}
\optCredit[.]{Writing - review \& editing}{\su}
}

% \section*{Data Availability}
% The data that support the findings of this study are available from the corresponding author upon reasonable request.

\appendix
\setcounter{equation}{0}
\renewcommand{\theequation}{A\arabic{equation}}

\section{Antenna Currents in Real Space}
\label{app:antCurReal}

\Cref{fig:antennaGeoMain} shows that the RF currents run in a straightforward way through the antenna. Taking the upper center $(z=0, \phi=2\pi)$ as our starting point, the full current $I_0$ will run diagonally downward across the upper helical strap. The current splits into upward and downward components at the left transverse strap. The current recombines at the second helical strap and runs diagonally upwards towards the right transverse strap. The current splits there and recombines at the first helical strap, thus completing the circuit.\\

It is clear that the transverse straps experience only azimuthal currents of value $\pm I_0/2$, depending on the azimuthal coordinate on the strap. Using the rectangle function $\rect{x}$ in \cref{eq:rect} we can describe a rectangle-shaped curve of width $w$, centered at a position $c$ along an axis coordinate $x$ as $\rect{(x-c)/w}$.\\

Along the $z$ coordinate the left and right transverse straps are located at $z_{L/R} = \mp(L - d_t)/2$. We can describe the current anywhere on the strap as $I_0/2$ and subtract $I_0$ on the central segment to account for reversed flow there. Lastly, we have to divide by the strap width to calculate the surface current density. We get the following description of the azimuthal current density on the transverse straps ($K_{\phi}^t$)

\begin{align}
\label{eqapp:Ktphi}
K_{\phi}^t =& \frac{I_0}{2 d_t}\left[\rect{\frac{\phi-\pi}{2\pi}} -2\rect{\frac{\phi-\pi}{\pi}}\right]\nonumber\\
&\times\left[\rect{\frac{z-z_L}{d_t}}+\rect{\frac{z-z_R}{d_t}}\right].
\end{align}

We can describe the helical strap currents by the same procedure. At $z=0$ the three helical segments are centered at $\phi_0 = 0, \pi$ and $2\pi$, respectively. The center line of each strap changes with the $z$ coordinate according to the strap pitch. Since the antenna is half-helical, each strap completes a $180\dg$ turn over an axial distance $L_h$. In right-handed antennas, the center line is moving up with increasing $z$. The opposite is true in left-handed antennas. The helicity-dependent strap pitch is then $\pi/L_h\times\psi$, where $\psi$ is the helicity according to \cref{eq:psi}. Each helical strap can therefore be described in dependence on the start position $\phi_0$ as 

\begin{align}
\label{eqapp:Hphi0}
H_{\phi_0}(\phi, z) & = \rect{\frac{z}{L_h}}\rect{\frac{\phi- \phi_0 -\frac{\pi \hel  z}{L_h}  }{\phi_w}}\rect{\frac{\phi-\pi}{2\pi}},
\end{align}

\noindent{}
where the last factor ensures that all straps are limited to the $(0, 2\pi)$ interval. The current magnitude on the helical strap is simply $I_0$ and can be split into azimuthal components ($I_{h,\phi}$) and axial components ($I_{h,z}$) as follows

\begin{align}
\frac{I_{h,z}}{L_h} &= \frac{I_{h,\phi}}{\pi R}\\
I_0^2 &= I_{h,\phi}^2 + I_{h,z}^2\\
\label{eqapp:Iz}
\Rightarrow |I_{h,z}| &= I_0\frac{1}{\sqrt{1+\gamma^2}}\\
\label{eqapp:Iphi}
\Rightarrow  |I_{h,\phi}| &= I_0\frac{\gamma}{\sqrt{1+\gamma^2}} \quad \text{with} \quad \gamma = \frac{\pi R}{L_h}.
\end{align}

\noindent{}
Another small calculation yields the angular strap width $\phi_w$

\begin{align}
\frac{\phi_w}{2\pi} &= \frac{d_{h,\phi}}{2\pi R}\\ 
\frac{d_h}{d_{h, \phi}} &= \cos(\alpha)\\
\tan(\alpha) &= \frac{\pi R}{L_h} = \gamma\\
\label{eqapp:phiW}
\Rightarrow \phi_w &= \sqrt{1+\gamma^2}\frac{d_h}{R}.
\end{align}

Combining \cref{eqapp:phiW,eqapp:Iphi,eqapp:Hphi0} yields the azimuthal current density on the helical straps ($K^{h}_\phi$) as 

\begin{align}
\label{eqapp:Khphi}
K^{h}_\phi(\phi, z) &= \frac{I_0 \gamma}{d_h\sqrt{1+\gamma^2}}\left(-H_{0}(\phi, z) + H_{\pi}(\phi, z) - H_{2\pi}(\phi, z)\right).
\end{align}

\Cref{eqapp:Iz} allows us to calculate the axial component of the current density directly from \cref{eqapp:Khphi}. If we were to draw the same diagram as \cref{fig:antennaGeoMain} for a left-handed antenna we would further find that the axial current directions on the helical straps are reversed. We therefore have for the axial current density on the helical straps ($K_z^h$)

\begin{align}
    \label{eqapp:Khz}
    K_z^h(\phi, z) &= \frac{\psi}{\gamma} K_\phi^h(\phi, z).
\end{align}

Combining \cref{eqapp:Ktphi,eqapp:Khphi,eqapp:Khz} gives us the full description of antenna surface currents 

\begin{align}
K_z =& \frac{I_0 \psi}{d_h \sqrt{1+\gamma^2}}\sum^{2}_{n=0} \sbr{ (-1)^{n+1}\rect{\frac{\phi- n\pi -\frac{\pi \hel  z}{L_h}  }{\phi_w}} }\nonumber\\
&\times\rect{\frac{\phi - \pi}{2\pi}}\rect{\frac{z}{L_h}}\\
K_{\phi} =& \frac{I_0}{2 d_t}\left[\rect{\frac{\phi-\pi}{2\pi}} -2\rect{\frac{\phi-\pi}{\pi}}\right]\nonumber\\
&\times\left[\rect{\frac{z-z_L}{d_t}}+\rect{\frac{z-z_R}{d_t}}\right] + \frac{\gamma}{\hel}K_z,
\end{align}

\noindent{}
as shown previously in \cref{eq:Kphi,eq:Kz}.

\renewcommand{\theequation}{B\arabic{equation}}
\section{Derivation of Antenna Currents in (m, z) Space}
\label{app:antCurMz}

We were able to express the surface current as a combination of box functions $\rect{\phi, z}$. Our first step is therefore to find the Fourier transform of such a function. The definition we use is the one shown previously in \cref{eq:mphiFtDef}, namely

\begin{align}
    \sft{f}(m) &= \frac{1}{2\pi} \int_{0}^{2\pi} f(\phi) \ftFac{m \phi} \,d\phi . 
\end{align}

For convenience, we will drop the $1/2\pi$ pre-factor until the end of this section. For a transform from a general $x$-space to $\nu$-space, defined as

\begin{align}
\label{eq:fourierBases}
g(\nu) &= \ft{f(x)} = \int^{\infty}_{-\infty} f(x) \ftFac{x \nu} \,dx,
\end{align}

\noindent{}
we have the following identities\cite{Kammler2000}

\begin{align}
\ft{\rect{x}} &= \ftSinc{\nu}\\
\ft{f\left(\frac{x}{a}\right)} &= |a|g\left(a \nu \right)\\
\label{eqapp:shift}
\ft{f(x-a)} &= \ftFac{a m} g(\nu),
\end{align}

\noindent{}
where we have used the definition $\sinc(\nu) = \sin(\pi \nu)/(\pi \nu)$. Together, they allow us to find the Fourier transform of a rectangle function with width $w$ and center $c$ as 

\begin{align}
\label{eqapp:boxTrans}
\ft{\rect{\frac{x-c}{w}}} &= w \ftFac{\nu c} \ftSinc{\nu w}.
\end{align}

Applying this result to the transformation of the transverse strap current in \cref{eqapp:Ktphi} from $\phi$-space to $m$-space we have

\begin{align}
    \ft{K_{\phi}^t} &=  \frac{I_0}{2 d_t}\ftFac{m \pi}\left[2\pi \ftSinc{2\pi m } - 2\pi \ftSinc{\pi m}\right]\\
    &= \frac{I_0}{d_t m} (-1)^m \left[\sin\left(m\pi \right) - {2}\sin\left(\frac{m \pi }{2}\right)\right]\\
    \label{eqapp:KtphimIntermediate}
    &= \frac{2 I_0}{m d_t} (-1)^{\frac{m-1}{2}} \quad\text{m odd},
\end{align}

\noindent{}
where we have omitted the axial box functions for brevity.\\

The helical straps require a transform of the $H_{\phi_0}$ function from  \cref{eqapp:Hphi0}. Using \cref{eqapp:boxTrans} and again omitting the axial box function we get 

\begin{align}
    \label{eqapp:helFt}
    \ft{H_{\phi_0}} & = \ftFac{m \phi_0}\phi_w \exp\left(-\frac{i m \pi \hel z}{L_h}\right)\ftSinc{m \phi_w}.
\end{align}

Importantly, we have assumed that the strap lies fully in the $(0,2\pi)$ interval. This is the case on the left in \cref{fig:antennaGeoMain} where we have straps with $\phi_0$ values of $\pi$ and $2\pi$. It is also the case on the right where we have straps with $\phi_0$ values of $\pi$ and $0$. In both regions application of \cref{eqapp:helFt} to \cref{eqapp:Khphi} yields the prefactor

\begin{align}
    \br{-e^{-2im\pi} + e^{-im\pi}}\phi_w\times \frac{ I_0 \gamma}{d_h \sqrt{1+\gamma^2}} &= -\frac{2\gamma I_0}{R} \quad\text{m odd},
\end{align}

\noindent{}
where we have used \cref{eqapp:phiW} to simplify the expression. The Fourier transform of the currents in the left and right parts of \cref{fig:antennaGeoMain} then becomes 

\begin{align}
\label{eqapp:KphimIntermediate}
    \sft{K}^h_\phi(m, z) &= -\frac{2\gamma I_0}{R} \ftSinc{m \phi_w} \exp\left(-\frac{i m \pi \hel z}{L_h}\right)\quad\text{m odd},
\end{align}

\noindent{}
again omitting the axial box function. We still have to analyze the middle part where we face the problem that the upper and lower strap are not fully in the $(0, 2\pi)$ interval. Defining $\alpha = \psi \pi/L_h$ and $\beta = \phi_w/2$, we know that the upper strap in this zone runs between $\alpha z - \beta$ and $2\pi$ and the lower strap between $0$ and $\alpha z + \beta$. We can perform the necessary integration directly

\begin{align}
    &\int_{0}^{2\pi}\pbr{  H_{2 \pi}(\phi, z) + H_{0}(\phi, z)} \ftFac{m\phi} \,d\phi\\
    &=  \int^{2 \pi}_{2 \pi + \alpha z - \beta} \ftFac{m \phi} \,d\phi + \int_{0}^{\alpha z + \beta} \ftFac{m \phi} \,d\phi\\
    &= -\frac{1}{im}\sbr{\expM{2} - \expM{2}\expA\expB[+] + \expA\expB - \expM{0}}\\
    &= \frac{\expA}{im}\sbr{\expB[+] - \expB}\\
    &= \frac{2 i \expA}{im}\sin\br{m \beta}\\
    &= \phi_w \exp\br{-\frac{i m \pi \hel z}{L_h}}\sinc\br{\frac{m \phi_w}{2\pi}}, 
\end{align}

\noindent{}
which is the same as \cref{eqapp:helFt} for $\phi_0 = 0$.
Using this result we get the following transform of \cref{eqapp:Khphi}

\begin{align}
\ft{K^{h}_\phi} & = \frac{I_0 \gamma}{d_h\sqrt{1+\gamma^2}}\ft{\left(-H_{0} + H_{\pi} - H_{2\pi}\right)}\\
&=    \frac{I_0 \gamma\br{-1 + \ftFac{m \pi}}}{d_h\sqrt{1+\gamma^2}}\phi_w \exp\br{-\frac{i m \pi \hel z}{L_h}}\sinc\br{\frac{m \phi_w}{2\pi}},
\end{align}

\noindent{}
which is identical to the solution in the left and right zones, given in \cref{eqapp:KphimIntermediate}.\\

By combining \cref{eqapp:KphimIntermediate,eqapp:KtphimIntermediate,eqapp:Khz} we get the full representation of antenna currents in $(m, z)$-space. Our original transform in \cref{eq:mphiFtDef} had a pre-factor of $1/2\pi$. Adding it and the axial box functions back in we arrive at

\begin{align}
\label{eqapp:KzTot}
\sft{K}_z &= -\frac{I_0 \hel}{\pi R}\ftSinc{m \phi_w} \exp\left(-\frac{i m \pi \hel z }{L_h}\right)\rect{\frac{z}{L_h}}\\
\label{eqapp:KphiTot}
\sft{K}_{\phi} &= \gamma \hel \sft{K}_z + \frac{I_0}{m \pi d_t} (-1)^{\frac{m-1}{2}} \left[\rect{\frac{z-z_L}{d_t}}+\rect{\frac{z-z_R}{d_t}}\right],
\end{align}

\noindent{}
as previously shown in \cref{eq:antCurZ,eq:antCurPhi}.

\renewcommand{\theequation}{C\arabic{equation}}
\section{Derivation of Antenna Currents in (m, k) Space}
\label{app:antCurMK}

\noindent{}
To transform from $z$-space to $k$-space we use the definition 

\begin{align}
    \dsft{f}(m, k) &= \frac{1}{2\pi} \int^{\infty}_{-\infty} \sft{f}(m, z) \ftFac{k z} \,dz,
\end{align}

\noindent{}
as shown previously in \cref{eq:mphiFtDef2}. We will again drop the $1/2\pi$ prefactor until the end of this section.\\

Using \cref{eqapp:boxTrans} it is straightforward to transform the contributions of the transverse strap currents in \cref{eqapp:KphiTot}.

\begin{align}
&\ft{\rect{\frac{z-z_L}{d_t}}+\rect{\frac{z-z_R}{d_t}}}\\
&= d_t \ftSinc{k d_t} \left( \ftFac{k z_L} + \ftFac{k z_R} \right)\\
&= d_t \ftSinc{k d_t} \left( \iftFac{k z_R} + \ftFac{k z_R} \right)\\
\label{eqapp:transTrick}
&= 2 d_t \ftSinc{k d_t} \cos\left(k z_R \right),
\end{align}

\noindent{}
where in the second step we have used that $z_{L} = -z_{R}$ as defined in \cref{eq:antCurSumEnd}.\\

We further have to transform the current on the helical straps in \cref{eqapp:KzTot}, where the following identity\cite{Kammler2000} will be useful

\begin{align}
\ft{f(x)e^{i a x}} &= g(\nu - a).
\end{align}

\noindent{}
The relevant transform then evaluates as follows

\begin{align}
&\ft{\exp\left(-\frac{i m \pi \hel z}{L_h}\right)\rect{\frac{z}{L_h}}}\\
&= L_h \left.\ftSinc{k L_h}\right|_{k \rightarrow k + \frac{m\pi\hel}{L_h}}\\
\label{eqapp:helTrick}
&= L_h \sinc\left[\frac{1}{2} \left(\frac{k L_h}{\pi} +m\hel\right)\right].
\end{align}

Applying \cref{eqapp:transTrick} to \cref{eqapp:KphiTot} and \cref{eqapp:helTrick} to \cref{eqapp:KzTot} and adding back the $1/2\pi$ prefactor directly yields the following current densities in $(m,k)$-space

\begin{align}
\dsft{K}_z &= - \frac{I_0 L_h \hel}{2 \pi^2 R} \ftSinc{m \phi_w} \sinc\left[\frac{1}{2} \left(\frac{k L_h}{\pi} + \hel m\right)\right]\\
\dsft{K}_\phi &= \gamma \hel \dsft{K}_z +\frac{I_0}{m \pi^2} (-1)^{\frac{m-1}{2}} \cos\left(k z_R \right) \ftSinc{k d_t},
\end{align}

\noindent{}
as previously shown in \cref{eq:Kzhmk,eq:Kphihmk}.

\section*{References}

\bibliography{references}\textbf{}
\end{document}
%
% ****** End of file aipsamp.tex ******